\documentclass[preprint,prd,aps,showpacs,preprintnumbers,amsmath,amssymb]{revtex4-1}
\usepackage{graphics,epsfig,subfigure}
\usepackage{diagbox}
\usepackage[usenames]{color}
\usepackage{color}
\usepackage{graphicx}
\usepackage{amsfonts}
\usepackage{indentfirst}
\usepackage{booktabs}
\usepackage{hyperref}
\hypersetup{hypertex=ture,backref=true,colorlinks=ture,linkcolor=blue,anchorcolor=blue,citecolor=blue}
\usepackage{float}
\usepackage{multirow}
\usepackage[section]{placeins}
\usepackage{array}
\usepackage{amsmath}

\begin{document}
\title{\Large \bf AdS Ellis wormholes with scalar field}
\author{Chen-Hao Hao}
\author{Xin Su}
\author{Yong-Qiang Wang\footnote{yqwang@lzu.edu.cn, corresponding author}}
\affiliation{ $^{1}$Lanzhou Center for Theoretical Physics, Key Laboratory of Theoretical Physics of Gansu Province,
	School of Physical Science and Technology, Lanzhou University, Lanzhou 730000, China\\
	$^{2}$Institute of Theoretical Physics $\&$ Research Center of Gravitation, Lanzhou University, Lanzhou 730000, China}

\begin{abstract}
In this paper, we study the spherically symmetric traversable wormholes with a scalar field supported by a phantom field in the anti-de Sitter (AdS) asymptotic spacetime. Despite coupling the scalar matter field, these wormholes remain massless and symmetric for reflection of the radial coordinate $r \rightarrow -r$. The solution possesses a finite Noether charge $Q$, which varies as a function of frequency $\omega$ with changes in the cosmological constant $\Lambda$ and the throat size $r_0$. Under specific conditions, an approximate ``event horizon'' will appear at the throat.
\end{abstract}

\maketitle

\section{INTRODUCTION}\label{Sec1}

The wormhole is a spacetime structure that can connect two different universes or distant regions in the same universe, it is featured by a minimal surface called ``throat''. Wormholes represent one of the most intriguing solutions in General Relativity (GR). In 1935, A. Einstein and his collaborator N. Rosen formally developed the wormhole theory, which was known as the ``Einstein-Rosen bridge''\cite{Einstein:1935tc}. We now understand that this solution is actually a part of the Kruskal extension of the Schwarzschild metric, and the ``Einstein-Rosen bridge'' is not traversable \cite{Kruskal:1959vx,Fuller:1962zza}. After that, the field fell silent for more than twenty years. Interest in this work was rekindled by J.A. Wheeler in the 1950s, he and C. W. Misner mentioned the word ``wormhole'' for the first time in 1957 \cite{Misner:1957mt}. To achieve traversable wormholes, a violation of the null energy condition seems necessary  \cite{Visser:1989kh}. The earliest traversable wormholes were discovered in \cite{Ellis:1973yv,Ellis:1979bh,Bronnikov:1973fh,Kodama:1978dw}, with further insights provided by Morris and Thorne in 1988 \cite{Morris:1988cz}. Among the many traversable wormhole solutions, the ``Ellis wormhole'' involves using a phantom field with an opposite sign in front of the kinetic energy term \cite{Lobo:2005us,Sushkov:2005kj,Lobo:2005yv,Bronnikov:2012ch,Kleihaus:2014dla}. Additionally, the more general ``Ellis-Bronnikov wormhole'' has been extensively studied \cite{Novikov:2009vn,Bronnikov:2013coa,Cremona:2018wkj,Huang:2020qmn}. Moreover, many traversable wormhole solutions do not require the exotic matter with negative energy density \cite{Bronnikov:2002rn,Kanti:2011jz,Maldacena:2020sxe,Blazquez-Salcedo:2020czn,Konoplya:2021hsm,Kain:2023pvp,Klinkhamer:2022rsj}. 

The previously mentioned solutions all exist in the background of asymptotically flat spacetime without cosmological constants. The advent of AdS/CFT correspondence \cite{Maldacena:1997re} has aroused great interest in the anti-de Sitter (AdS) spacetime with negative cosmological constants. Particularly within this framework, the ER=EPR \cite{Maldacena:2013xja} proposal has underscored the importance of studying wormhole solutions in AdS spacetime  \cite{Gao:2016bin,Maldacena:2017axo,vanBreukelen:2017dul,Maldacena:2018lmt,Dai:2020ffw,Bintanja:2021xfs,Kundu:2021nwp,Kain:2023ore}. However, the exact wormhole solutions are elusive in AdS spacetime, contrary to the asymptotically flat case, and so far, not so many fully-fledged examples of AdS wormhole solutions are available in the literature. Notable examples include solutions obtained via the cut-and-paste technique \cite{Lemos:2003jb,Lemos:2004vs}, the modified gravity solution \cite{Maeda:2008nz}, the dynamical solution \cite{Maeda:2012fr}, the Ricci-flat/AdS correspondence solution \cite{Wu:2022gpm}, the Ads solution generated from flat spacetime \cite{Nozawa:2020gzz}, and the asymptotically locally AdS solution \cite{Anabalon:2018rzq}. Moreover, intriguing investigations into the properties of AdS wormholes have been conducted \cite{Korolev:2014hwa,Franciolini:2018aad,Chatzifotis:2020oqr,Blazquez-Salcedo:2020nsa}. It is worth noting that in \cite{Blazquez-Salcedo:2020nsa}, the AdS asymptotic Ellis wormhole was successfully constructed, and the properties of this solution were explored through numerical methods.

On the other hand, the attempts to combine the matter fields with GR to find an exact solution can be traced back to 1955. J. A. Wheeler got an unstable solution by coupling the classical fields of electromagnetism with general relativity and he named these objects ``geons''\cite{Wheeler:1955zz,Power:1957zz}. Later, Kaup et al. obtained Klein-Gordon geons (i.e., boson stars) by replacing the massless vector field with a massive complex scalar field \cite{Kaup:1968zz}. Ruffini also independently studied boson stars by considering quantized real scalar fields \cite{Ruffini:1969qy}. And now, the enthusiasm for the study of Boson stars has grown day by day, except that the original boson stars are spherically symmetric and composed of free scalar fields with fundamental configurations. It is generalized to the cases of rotation \cite{Schunck:1996he,Yoshida:1997qf}, the excited BSs \cite{Bernal:2009zy,Collodel:2017biu,Wang:2018xhw}, static multipolar BSs \cite{Herdeiro:2021mol} and the AdS asymptotic BSs \cite{Astefanesei:2003qy,Buchel:2013uba,Maliborski:2013ula,Fodor:2015eia,Brihaye:2013hx,Liu:2020uaz}. Recently, in \cite{Dzhunushaliev:2014bya,Hoffmann:2017jfs,Yue:2023ela,Ding:2023syj,Hao:2023igi,Su:2023xxk}, the solutions of Ellis wormholes coupling scalar field, Proca fields, and Dirac fields were constructed respectively. These solutions not only reveal how nontrivial topological spacetime impacts the physical properties of the matter fields but also demonstrate intriguing ``extreme'' behavior under a specific parameter range. In this study, we extend the Ellis wormhole solution with a scalar field to AdS spacetime, investigating its geometric structure and properties. The solution possesses a finite Noether charge $Q$, which varies as a function of frequency $\omega$ with changes in the cosmological constant $\Lambda$ and the throat size $r_0$. Under specific conditions, an approximate ``event horizon'' will appear at the throat.

The paper is organized as follows. In Sec.~\ref{sec2}, we present the model of four-dimensional
Einstein's gravity coupled to a phantom field and a scalar field in the anti-de Sitter (AdS) asymptotic spacetime. In Sec.~\ref{sec3}, the boundary conditions are studied. We perform the series expansion for the equations of the metric and the scalar field to study the asymptotic behavior. The numerical results of the solution are discussed in Sec.~\ref{sec4}. We conclude in Sec.~\ref{sec5} with a summary and illustrate the range for future work.

\section{THE MODEL}\label{sec2}

\subsection{Action}

We consider the Einstein-Hilbert action including the Lagrangian for two massive Dirac fields and the phantom scalar field, the action is given by
\begin{equation}\label{action}
 S=\int\sqrt{-g}d^4x\left(\frac{R}{2\kappa}+\mathcal{L}_{p}+\mathcal{L}_{s}\right),
\end{equation}
where $R$ is the Ricci scalar.  The term
$\mathcal{L}_{p}$ and $\mathcal{L}_{s}$ are the Lagrangians defined by with

\begin{eqnarray}
\mathcal{L}_{s}  &= & -\nabla_a\Psi^*\nabla^a\Psi  - \mu_0^2\Psi\Psi^*,  \\
\mathcal{L}_{p}  & =   &   \nabla_a\Phi\nabla^a\Phi \ .
\end{eqnarray}

Here $\Psi$  and $\Phi$ represent the complex scalar field and the phantom field, respectively.
By varying the action (\ref{action}) with respect to the metric, we can obtain the Einstein equations
\begin{equation}
  \label{eq:EKG1}
R_{\mu\nu}-\frac{1}{2}g_{\mu\nu}R-\kappa T_{\mu\nu}+ \Lambda g_{\mu\nu}=0 \ ,
\end{equation}
with stress-energy tensor
\begin{equation}
T_{\mu\nu} = g_{\mu\nu}({{\cal L}}_s+{{\cal L}}_p)
-2 \frac{\partial ({{\cal L}}_s+{{\cal L}}_p)}{\partial g^{\mu\nu}} \ ,
\end{equation}
and the matter field equations by varying with respect to the phantom field and Dirac fields.
\begin{equation}
  \label{eq:EKG2}
  \Box\Psi-\mu_0^2\Psi=0,
\end{equation}
and
\begin{equation}
  \label{eq:EKG2}
  \Box\Phi=0.
\end{equation}

\subsection{Ansatze}

We consider the  general static spherically symmetric solution with a wormhole,
and adopt the Ansatzes as follows, see {\em e.g.}~\cite{Blazquez-Salcedo:2020nsa}:
\begin{equation}  \label{line_element1}
 ds^2 = -F(r) N(r) dt^2 + \frac{p(r)}{F(r)}   \left[ \frac{1}{N(r)}d r^2 + h(r) (d \theta^2+\sin^2 \theta d\varphi^2)   \right]\,,
\end{equation}
here $N(r) = 1 - \frac{\Lambda r^2}{3}$,  $h(r)=r^2+r_0^2$ with  the throat parameter  $r_0$,
 and $r$  ranges from positive infinity to negative infinity.
It should be emphasized that the two limits $r\rightarrow \pm\infty$ correspond to two distinct asymptotically flat spacetime and in the pure Einstein gravity ($\Lambda = 0$), the above metric describes a static symmetric Ellis wormhole with scalar field \cite{Dzhunushaliev:2014bya,Ding:2023syj}.
Furthermore, we assume stationary complex scalar  field and phantom field in the form
\begin{eqnarray}  \label{an2}
  \Psi&=\psi(r)e^{i\omega t}, \;\;\;\;  \Phi&=\phi(r).
\end{eqnarray}
Here, $\psi$ is only a real function of the radial coordinate $r$, and the constant $\omega$ is referred to as the synchronization frequency. Moreover, the phantom field $\Phi$ is also a real function and is independent of the time coordinate $t$.

Variation of the action with respect to the
complex scalar field and the phantom field
leads to the equations

\begin{equation}
\begin{split}
2 F^{2} N^{2} p h^{\prime} \psi {\prime}+h\left(2(\omega ^{2}-\mu_0 ^{2} FN) p^{2} \psi +F^{2} N^{2} p^{\prime} \psi ^{\prime}+2F^2Np\left(N^{\prime}\psi ^{\prime}+N\psi^{\prime\prime}\right)\right) =0,
\end{split}
\end{equation}

\begin{equation}
(\phi'h N \sqrt{p})' = 0 \ .
\end{equation}

Substituting the above Ansatzes into the Einstein equations leads to the following field equations

\begin{equation}
\begin{split}
& -2 h p(4\kappa \omega ^{2} p \psi^{2}+N^{2} F^{\prime 2})+F N\bigl(2 N p F^{\prime} h^{\prime}+h(4 p^{2}(\Lambda+ \kappa \mu_{0}^{2}  \psi^{2})+N F^{\prime} p^{\prime} \\& +  2 p(F^{\prime} N^{\prime}+N F^{\prime \prime}))\bigr)+F^{2} N\left(N^{\prime}\left(2 p h^{\prime}+h p^{\prime}\right)+2 h p N^{\prime \prime}\right) =0,
\end{split}
\end{equation}

\begin{equation}
\begin{split}
&-8 \kappa \omega^{2} h p^{3} \psi^{2}+8 F h N p^{3}\left(\Lambda+\kappa \mu_{0}^{2} \psi^{2}\right)+F^{2} N\bigl(-h N p^{\prime 2} \\ & + 2 p^{2}(-2+2 h^{\prime} N^{\prime}+N h^{\prime \prime}+h N^{\prime \prime})+p(3(N h^{\prime}+h N^{\prime}) p^{\prime}+2 h N p^{\prime \prime})\bigr) =0,
\end{split}
\end{equation}

\begin{equation}
\begin{split}
&-h^{2} p^{2}\left(4 \kappa \omega^{2} p \psi^{2}+N^{2} F^{\prime 2}\right)+2 F h^{2} N p^{2}\left(2 p\left(\Lambda+\kappa \mu_0^{2} \psi^{2}\right)-F^{\prime} N^{\prime}\right)+ \\ & F^{2} N\left(N p^{2} h^{\prime 2}+2 h p\left(p\left(-2+h^{\prime} N^{\prime}\right)+N h^{\prime} p^{\prime}\right)+h^{2}\left(2 p N^{\prime} p^{\prime}+N p^{\prime 2}+2 \kappa N p^{2}\left(\phi^{\prime 2}-2 \psi^{2}\right)\right)\right) =0.
\end{split}
\end{equation}

These five equations are divided into three groups: three of these equations (10), (12), and (13) are solved together, by solving these OED equations numerically, we can get all information about metric functions $F(r)$ and $p(r)$, field $\psi(r)$. The remaining one equation (14) is
treated as the constraint and used to check the numerical accuracy of the method.  Furthermore, as the derivative in Eq. (11) happens to be
zero, we can transform the last expression into the following form
\begin{equation}
\phi' = \frac{\sqrt{\cal D}}{h N \sqrt{p}}\ ,
\end{equation}
the $\cal D$ is a constant that represents the scalar charge of the phantom field and can be used to check the accuracy of numerical calculations. Its value as a function of frequency $\omega$ should be the same at different locations while fixing $r_0$ or $\Lambda$.
We give the expression of scalar charge $\cal D$ by taking the above Eq. (15) into the Eq. (14)

\begin{eqnarray}
\begin{split}
&{\cal D}
  = h^{2} p^{2}\left(4 \kappa \omega^{2} p \psi^{2}+N^{2} F^{\prime 2}\right)+2 F h^{2} N p^{2}\left(-2 p\left(\Lambda+\kappa \mu_0^{2} \psi^{2}\right)+F^{\prime} N^{\prime}\right) \\ & + F^{2} N\left(-N p^{2} h^{\prime 2}-2 h p\left(p\left(-2+h^{\prime} N^{\prime}\right)+N h^{\prime} p^{\prime}\right)+h^{2}\left(-p^{\prime}\left(2 p N+N p^{\prime}\right)+4 \kappa N p^{2} \psi^{2}\right)\right).
\end{split}
\end{eqnarray}

\section{BOUNDARY CONDITIONS}\label{sec3}

Before numerically solving the differential equations instead of seeking the analytical
solutions, we should obtain the asymptotic behaviors of the three functions $\psi$, $F$, $p$
which is equivalent to giving the boundary conditions we need.

For our work, the solution is symmetric and continuous at the origin, and we can get the solution in the whole spacetime at once, which means there is no need to limit the boundary conditions at the origin. To study the asymptotic behavior of the metric functions in the limit $r \rightarrow \infty$, we perform the series expansion for Eqs. (12) and (13) to obtain the asymptotic expansion for the functions

\begin{equation}
F(r)=F_{\infty} + F_{\infty } \frac{r _0^{2}}{3 r^{2}}-\frac{F_{\infty }(15-48 r _0^{2}+4 r _0^{4} \Lambda)}{60 \Lambda r^{4}}+o\left(r^{-6}\right),
\end{equation}

\begin{equation}
p(r)=p_{\infty} - p_{\infty } \frac{r _0^{2}}{3 r^{2}}+\frac{p_{\infty }(-45+ 108 r _0^{2}+ 56 r _0^{4} \Lambda)}{60 \Lambda r^{4}}+o\left(r^{-6}\right).
\end{equation}

Obviously, the odd terms vanish identically and we can get the large-$r$ expansions of $g_{tt}$ for this wormhole solution
\begin{equation}
\begin{split}
-\left.g_{t t}\right|_{r \rightarrow \infty}= & -\frac{\Lambda F_{\infty} r^{2}}{3}+F_{\infty}\left(1-\frac{\Lambda r_{0}^{2}}{9}\right) \\
& +\frac{F_{\infty} r_{0}^{2}}{3 r^{2}}\left(1+\frac{\frac{5}{r_0^2}-16+\frac{4}{3}r_0^2 \Lambda}{20}\right)+O\left(\eta^{-4}\right).
\end{split} 
\end{equation}

The odd terms vanish, and the metric functions have the same asymptotic behaviors at $r \rightarrow -\infty$ and $r \rightarrow \infty$, they all return to the Minkowski spacetime. The appropriate boundary conditions to be imposed on the metric functions and scalar field function at infinity are given by
\begin{equation}
F( \pm \infty)=p( \pm \infty)=1,
\end{equation}
\begin{equation}
\psi( \pm \infty)=0.
\end{equation}

Meanwhile, considering the expression for the mass of the wormholes, obtained from the ADM formalism, the vanishing of the $1/r$ term also
implies that the mass of these symmetric wormholes vanishes, this has not changed with the addition of the scalar matter field. Interestingly, when the cosmological constant $\Lambda$ is 0, this solution degenerates back to the Ellis wormhole with a scalar field and has finite ADM mass, which is encoded in the asymptotic expansion of metric components (which means odd terms reappear)
\begin{eqnarray}
g_{tt}= -1+\frac{2 M}{r}+\cdots \ .
\end{eqnarray}

On the other hand, the action of the complex scalar field is invariant under the $U(1)$ transformation $\psi\rightarrow e^{i\alpha}\psi$ with a constant $\alpha$. According to Noether's theorem, there is a conserved current corresponding to the complex scalar field:

\begin{equation}\label{equ9}
 J^{\mu} = -i\left(\psi^*\partial^\mu\psi - \psi\partial^\mu\psi^*\right), \;\;\;\;\;\;\; J^\mu_{\,\,\,; \mu} =0 \;.
\end{equation}

Integrating the timelike component of the above-conserved currents on a spacelike hypersurface $\cal{S}$, one could obtain the Noether charge in the symmetric case:
\begin{eqnarray}
Q  &=& \int_{\cal S}J_s^t \nonumber \\
&= &- \int J^t \left| g \right|^{1/2} dr d\Omega_{2}.
\end{eqnarray}

\section{NUMERICAL RESULTS}\label{sec4}

In this work, all the numbers are dimensionless as follows
\begin{eqnarray}
r \rightarrow r\mu \hspace{5pt}, \hspace{5pt} \phi \rightarrow \phi \kappa^{-1/2}\hspace{5pt}, \hspace{5pt} \omega \rightarrow \omega/\mu \hspace{5pt}.
\end{eqnarray}
Without loss of generality, we can fix the specific parameters as $\mu_0 = 1$ and $\kappa= 2$.
To facilitate numerical calculations, we transform the radial coordinates by the following equation
\begin{eqnarray}
\label{transform}
x= \frac{2}{\pi}\arctan(r) \;,
\end{eqnarray}
map the infinite region ($-\infty$,$+\infty$) to the finite region (-1,1).
This allows the ordinary differential equations to be approximated by algebraic equations. The grid with 2000 points covers the integration region and the relative errors are less $10^{-5}$.

We focus on two variable parameters: the cosmological constant $\Lambda$ and the throat size $r_0$. In the subsequent presentation of results, we will keep one of these parameters constant while varying the other to investigate solution properties. Furthermore, the symmetry of the solution means that the throat is located at the center of the wormhole, at $x = 0$. Without loss of generality, we only show the results for the cosmological constant $\Lambda$ from 0 to -10, it can take any value less than 0.

\subsection{The cosmological constant $\Lambda = 0$}
\begin{figure}[H]
\begin{center}
\subfigure{\includegraphics[width=0.45\textwidth]{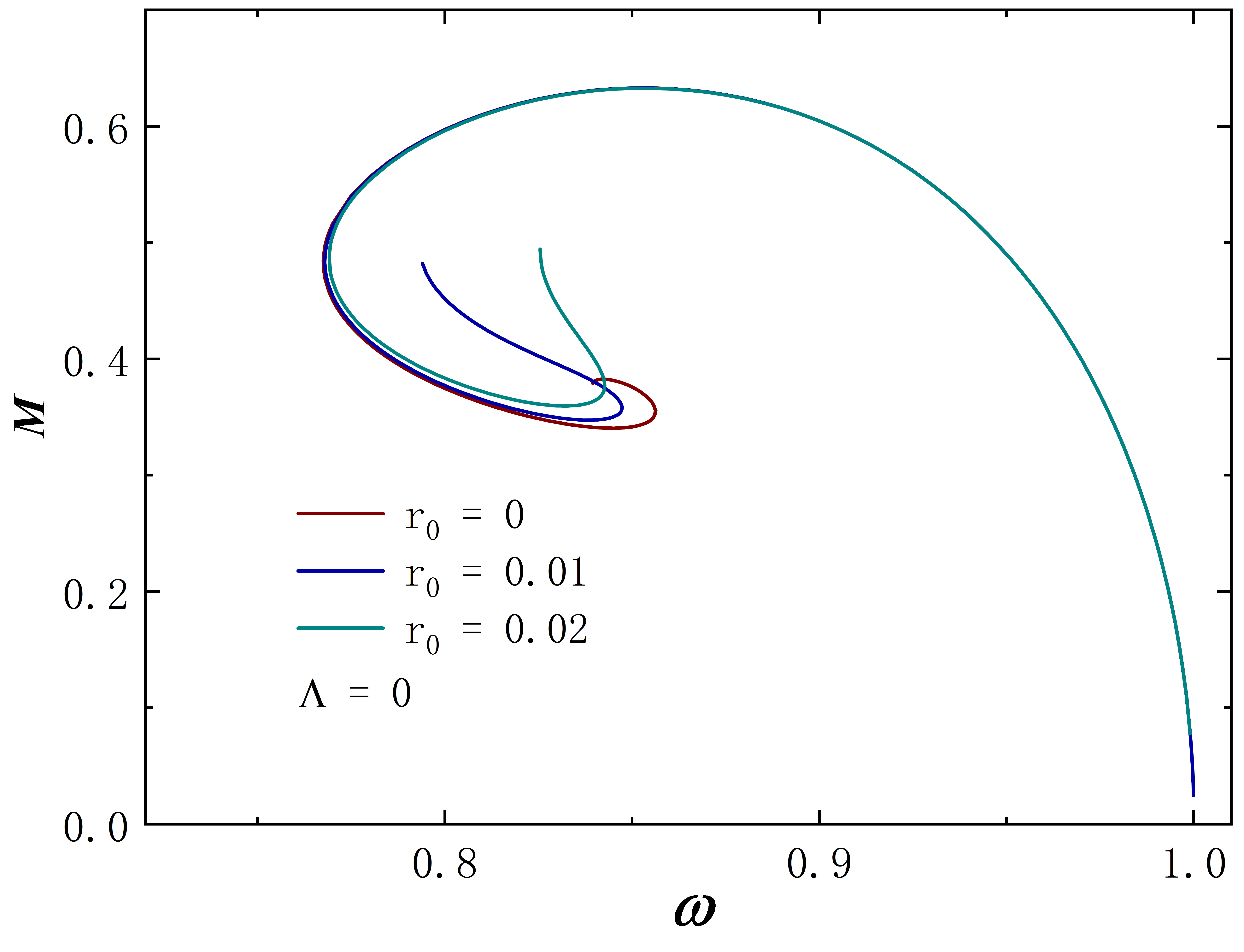}}
\subfigure{\includegraphics[width=0.45\textwidth]{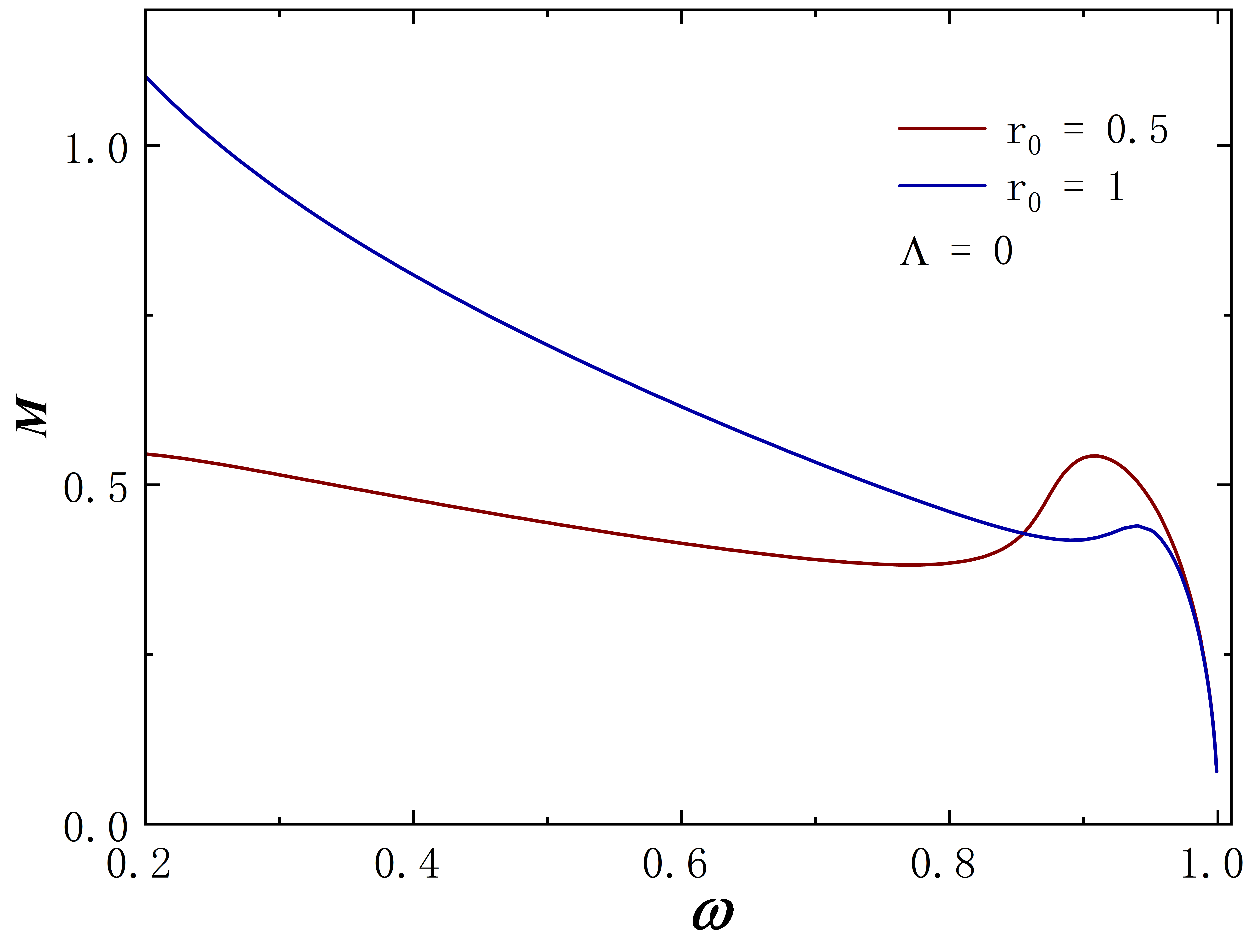}}
\subfigure{\includegraphics[width=0.45\textwidth]{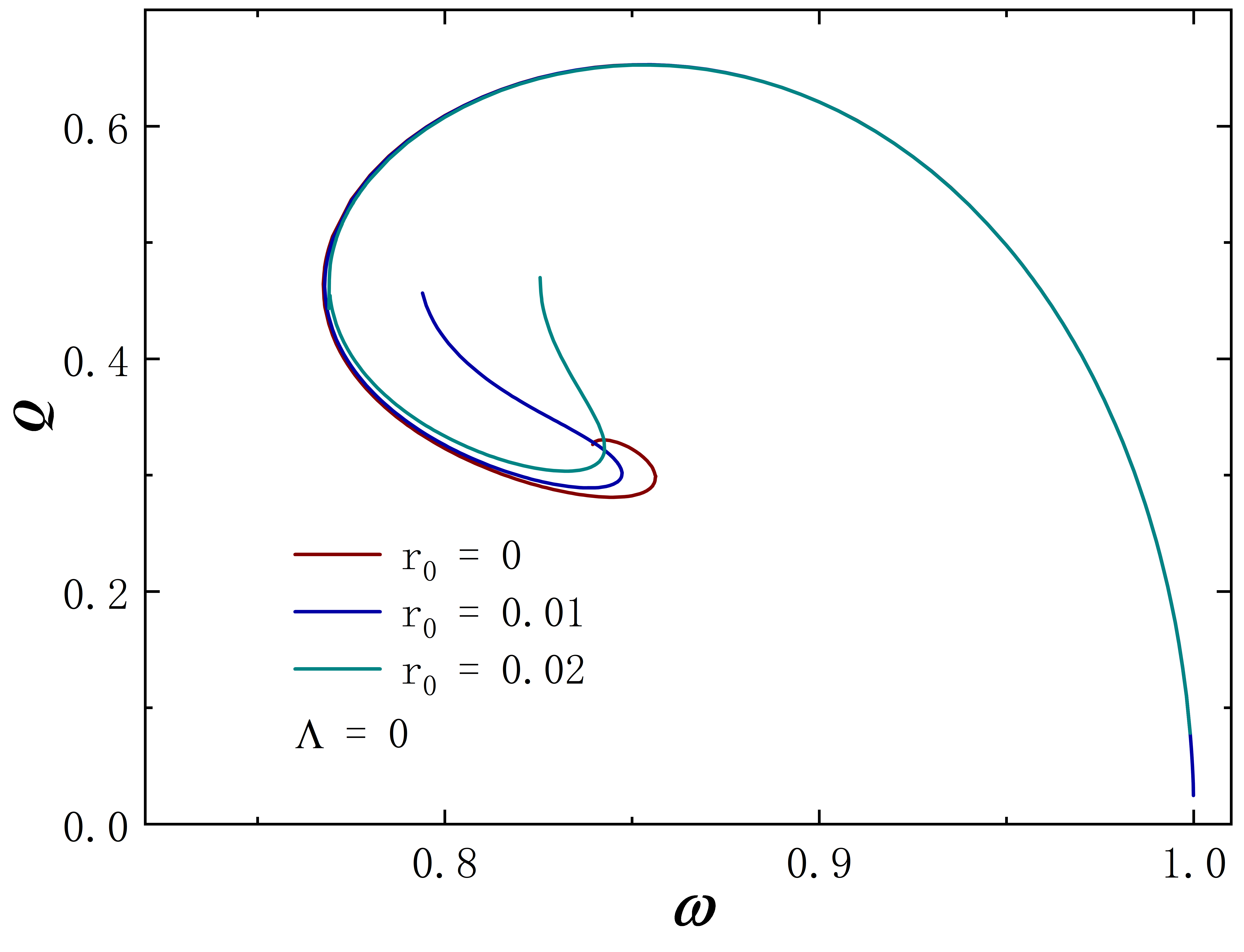}}
\subfigure{\includegraphics[width=0.45\textwidth]{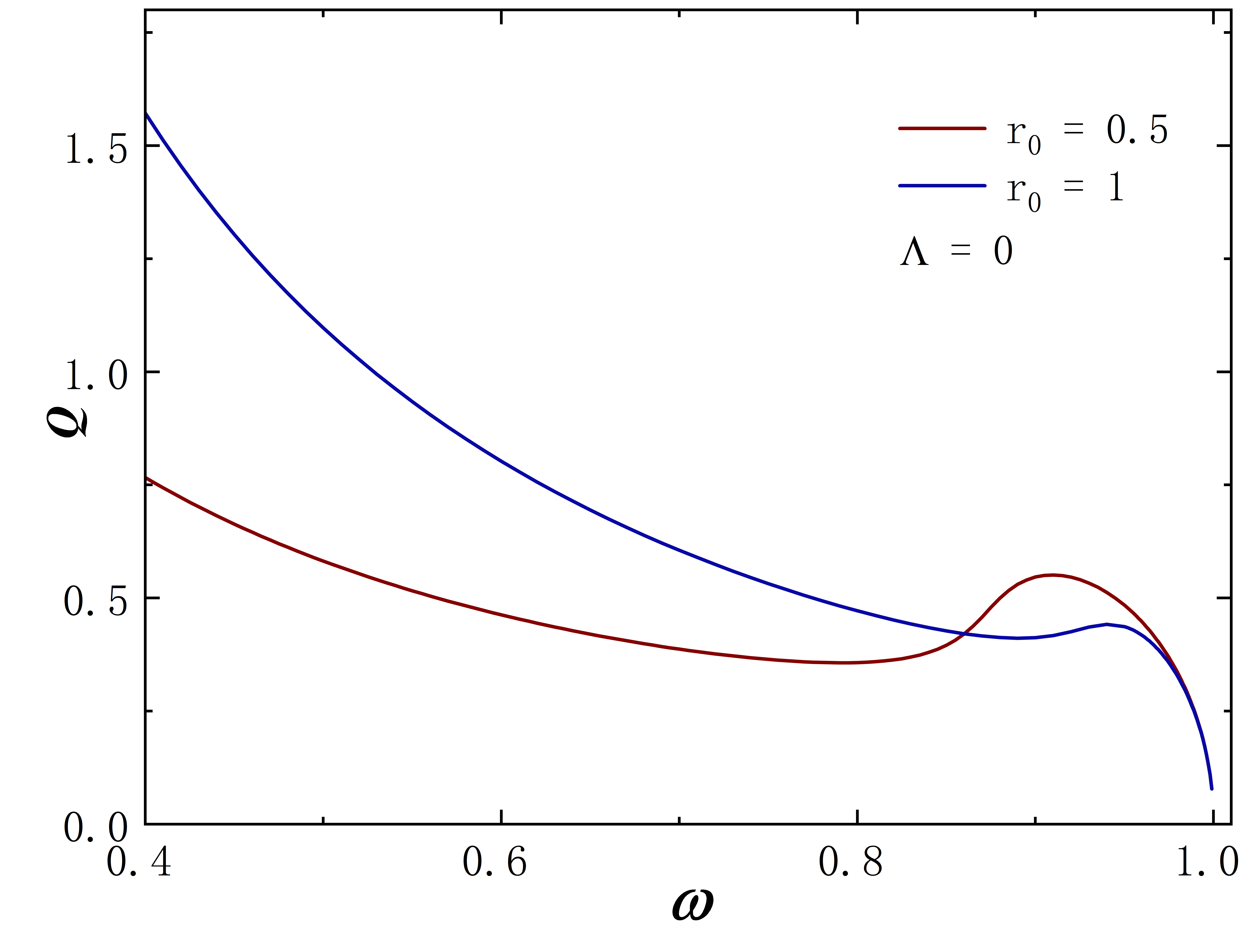}}
\end{center}
\caption{The ADM mass $M$ and Noether charge $Q$ as the function of frequency $\omega$ for some values of $r_0$ with $\Lambda = 0$.}
\label{phase1}
\end{figure}

When the cosmological constant $\Lambda = 0$, the solution degenerates to the static symmetric Ellis wormhole with a scalar field. At this juncture, we show the ADM mass $M$ and Noether charge $Q$ for this solution in Fig.\ref{phase1}, and the numerical results are aligned with the \cite{Ding:2023syj}. When $r_0$ is small, the solution will be limited in a tight range of $\omega$, and the curve shapes are just like the well-known spiral curve of Bonson stars. As $r_0$ increases, the spiral curve configuration gradually unfolds, eventually resulting in a single branch, such as the case of $r_0 = 1$. Notably, the mass $M$ and charge $Q$ at this time will sharply grow almost linearly as $\omega$ decreases, so we do not show the data when $\omega$ is very small. Further properties of this solution without cosmological constants will not be presented. Interested readers are encouraged to refer to the \cite{Ding:2023syj} for additional details.

\subsection{The cosmological constant $\Lambda < 0$}

In the context of a negative cosmological constant, the odd terms in the metric $g_{tt}$ expansion vanish, implying that the ADM mass of the solution becomes zero. To investigate the solution's properties, we study the functional relationship between the Noether charge $Q$ and frequency $\omega$ under varying cosmological constant $\Lambda$ for three distinct throat size $r_0$ groups, ranging from small to large.

\begin{figure}[H]
\begin{center}
\subfigure{\includegraphics[width=0.45\textwidth]{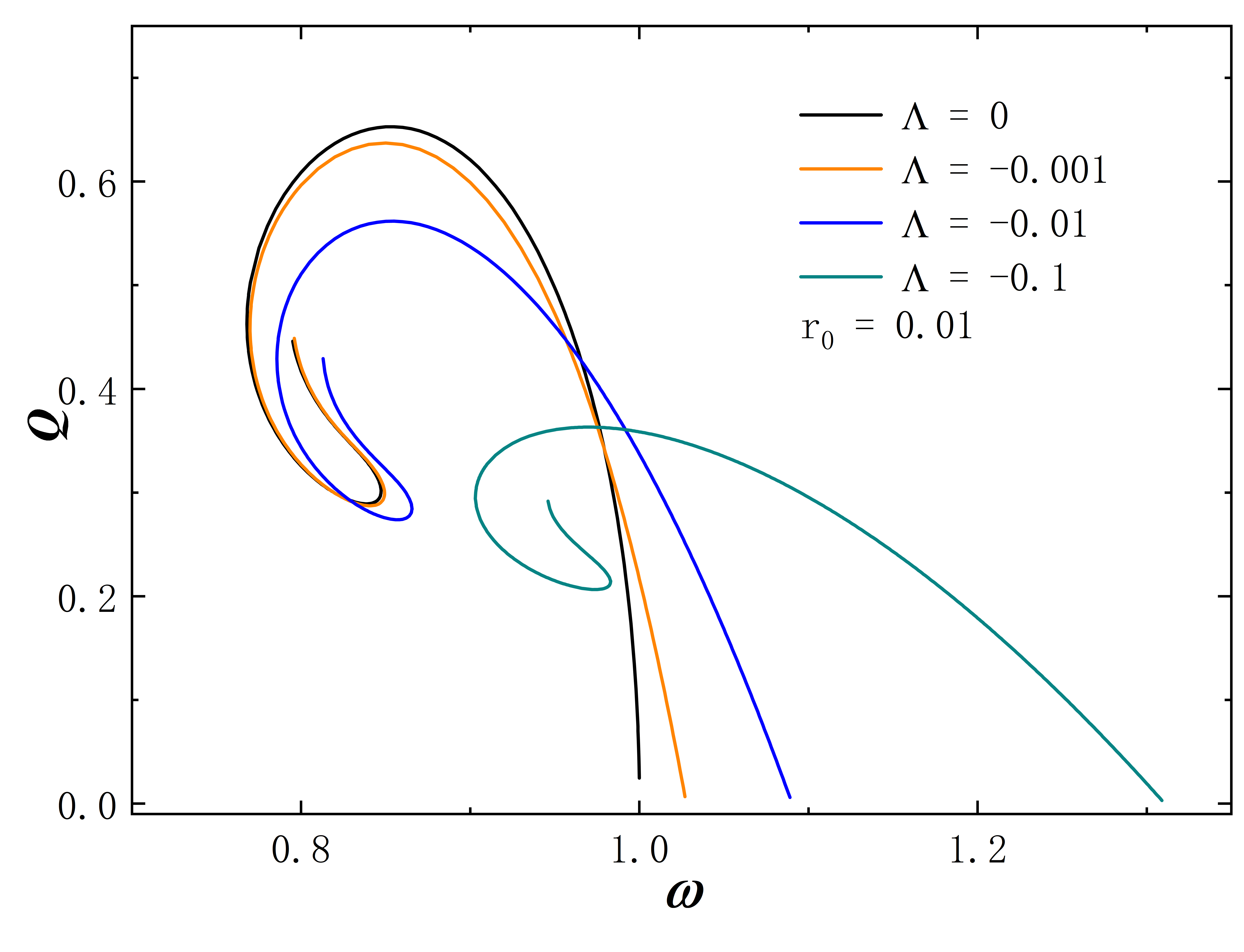}}
\subfigure{\includegraphics[width=0.45\textwidth]{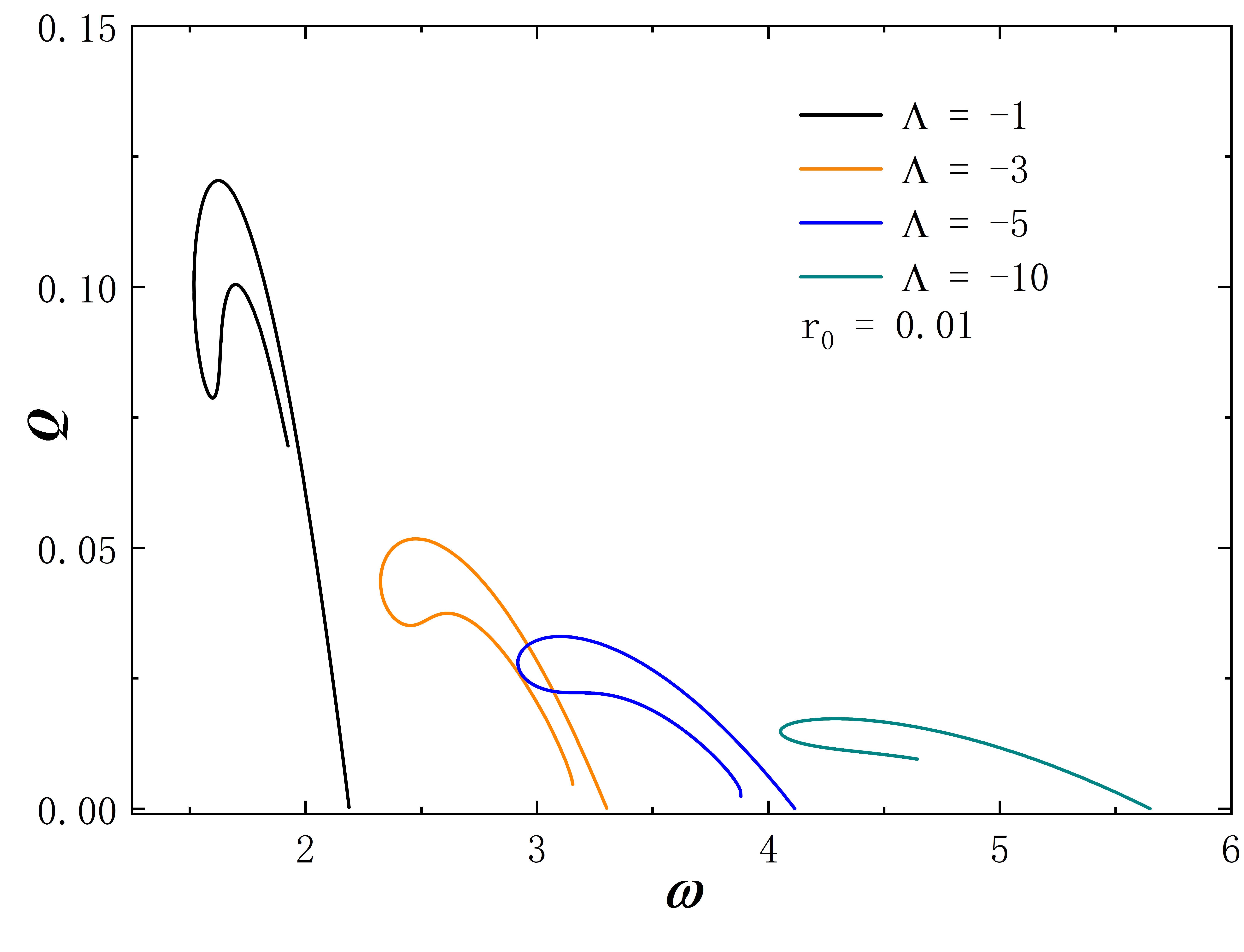}}
\subfigure{\includegraphics[width=0.45\textwidth]{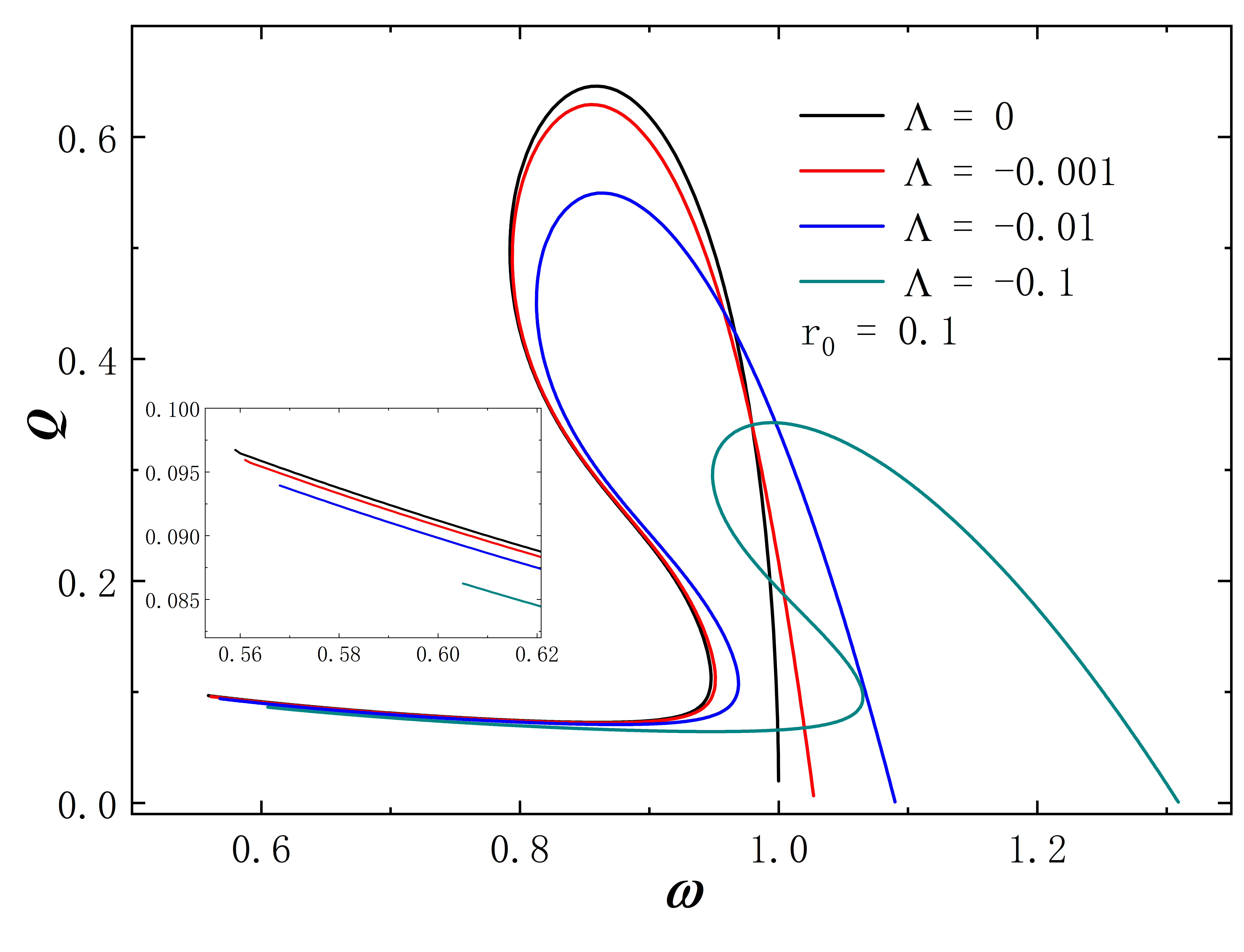}}
\subfigure{\includegraphics[width=0.45\textwidth]{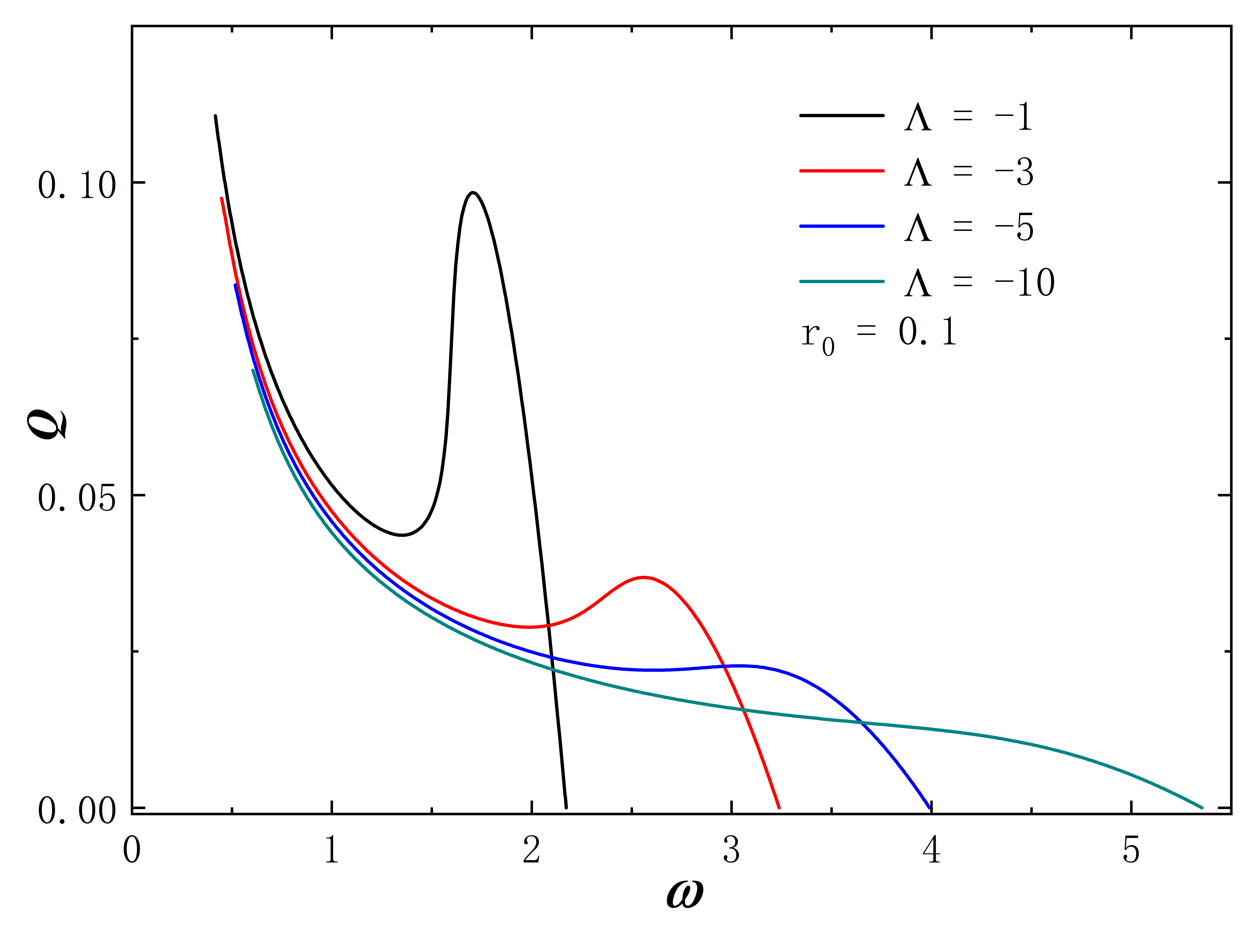}}
\subfigure{\includegraphics[width=0.45\textwidth]{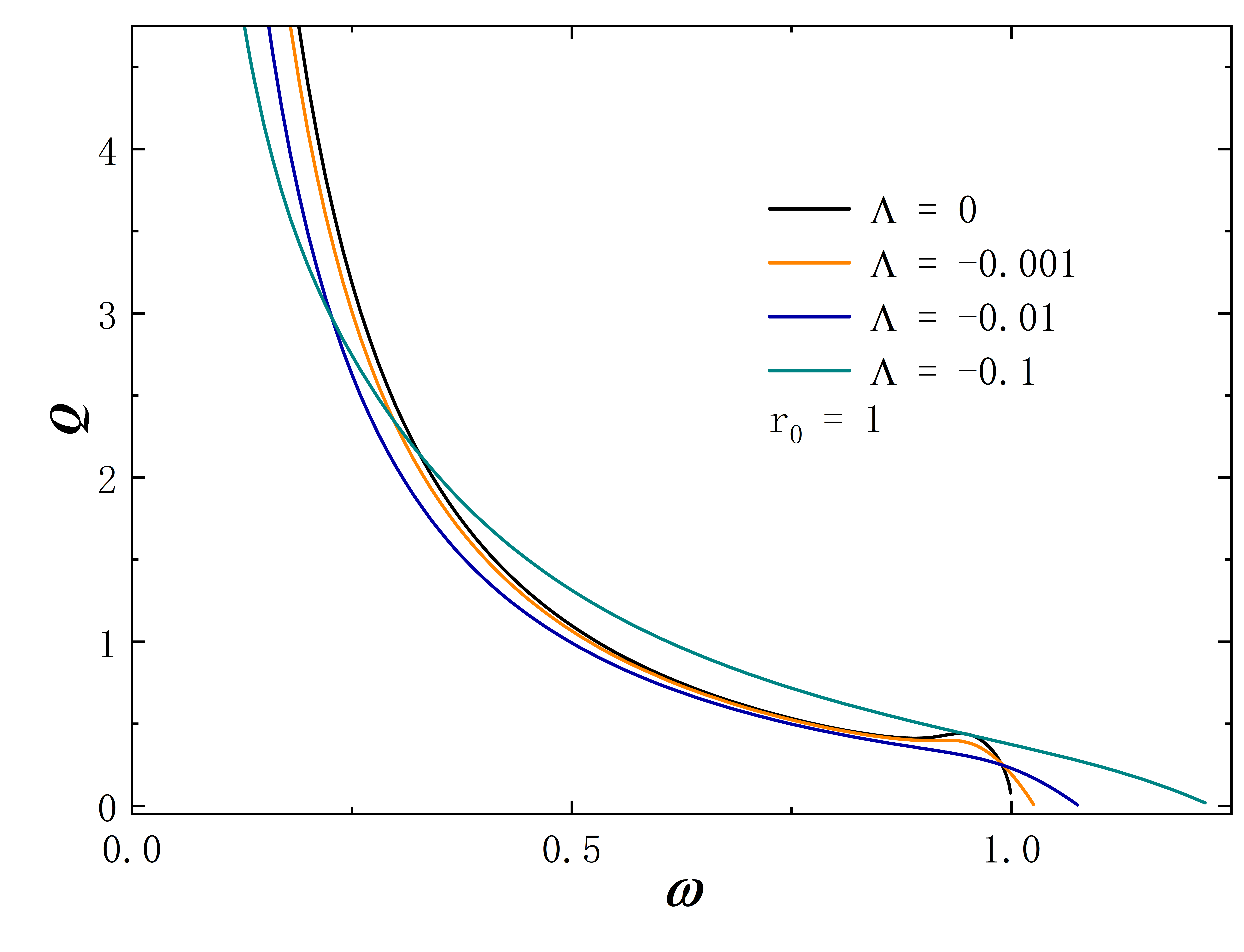}}
\subfigure{\includegraphics[width=0.45\textwidth]{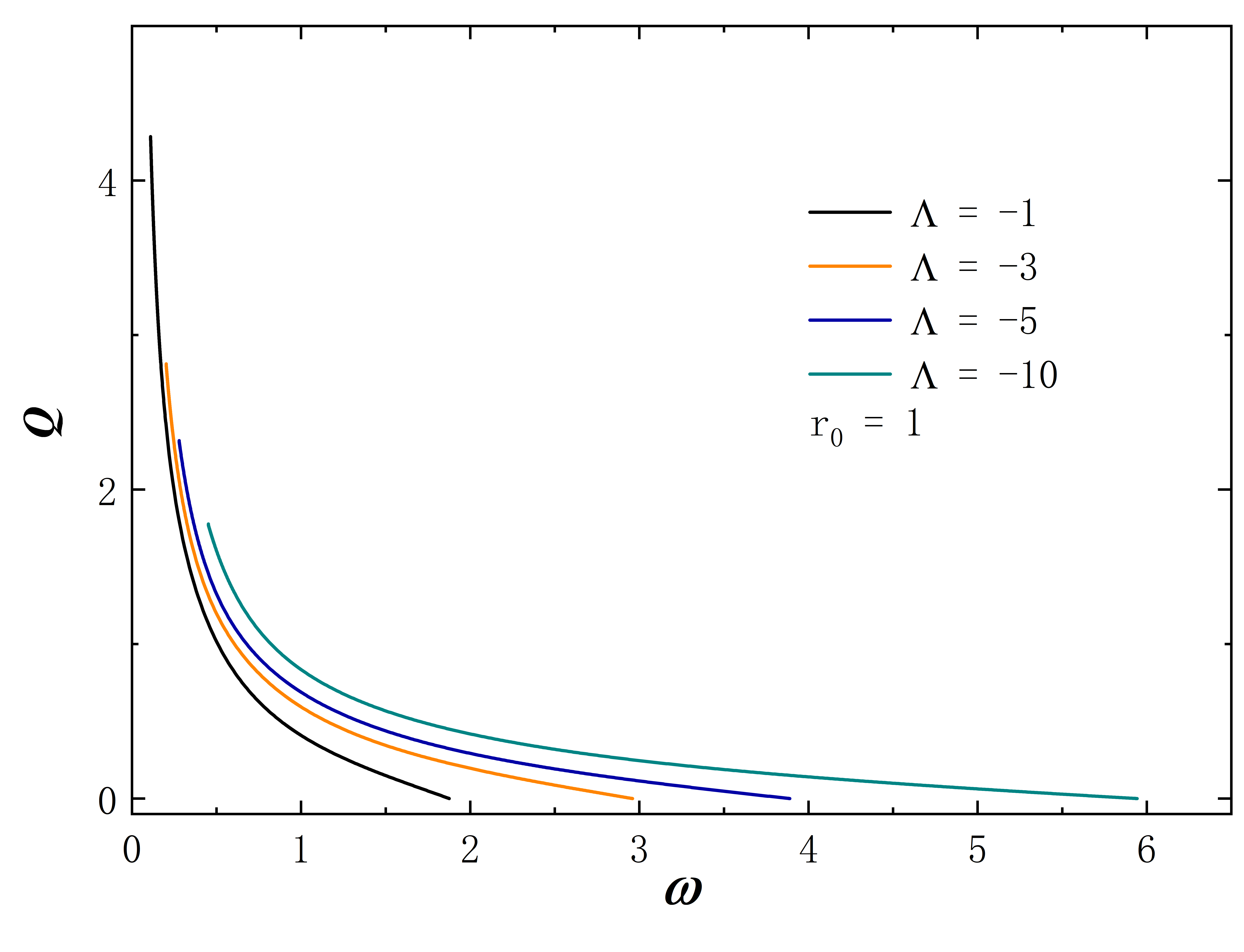}}
\end{center}
\caption{The Noether charge $Q$ as a function of frequency $\omega$ under varying cosmological constant $\Lambda$ for three distinct throat size $r_0$ groups. ($r_0 = 0.01, 0.1, 1$).}
\label{phase2}
\end{figure}

The Noether charge $Q$ as a function of frequency $\omega$ is shown in Fig.\ref{phase2}. When the cosmological constant approaches zero, the solutions exhibit minimal deviation from $\Lambda = 0$. This behavior is evident both in the curve shape and the magnitude of the Noether charge $Q$. As $\Lambda$ decreases further, gradual changes in solution properties become apparent. Specifically, for $r_0 = 0.01$, the reduction in the $\Lambda$ leads to fewer solution branches, the gradual unfolding of the corresponding spiral curve, and a decrease in the value of $Q$. The situation of $r_0 = 0.1$ exhibits similar behavior. However, for larger throat sizes, such as $r_0 = 1$, the situation diverges. At this time, the solution exhibits no branches regardless of the cosmological constant's value and the Noether charge $Q$ does not exhibit an obvious monotonic change as $\Lambda$ decreases. Furthermore, the presence of a negative cosmological constant alters the solution's existence domain, which is manifested in that as $\Lambda$ decreases, the frequency range of the solution expands and gradually shifts to the right.

\begin{figure}
\begin{center}
\subfigure{\includegraphics[width=0.45\textwidth]{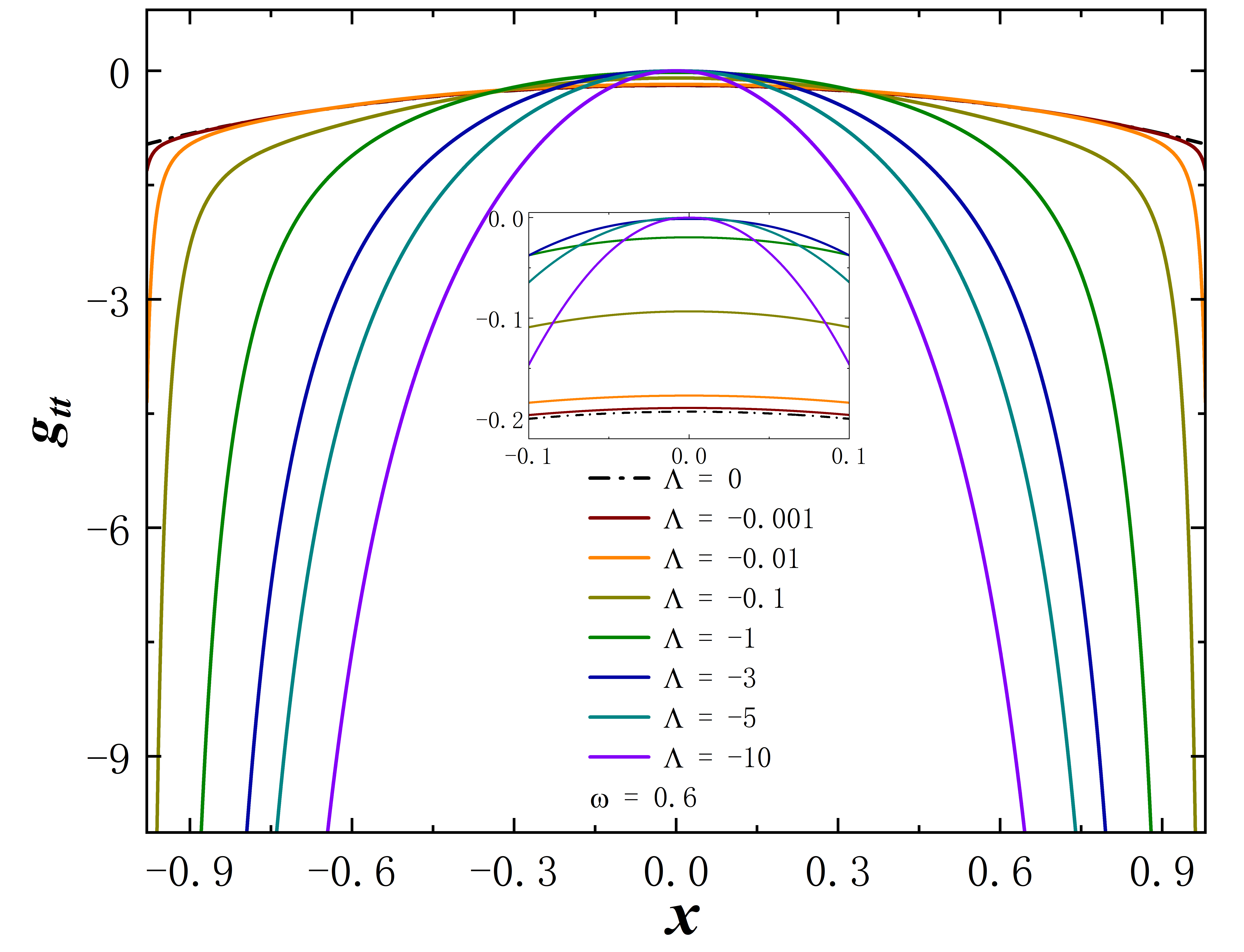}}
\subfigure{\includegraphics[width=0.45\textwidth]{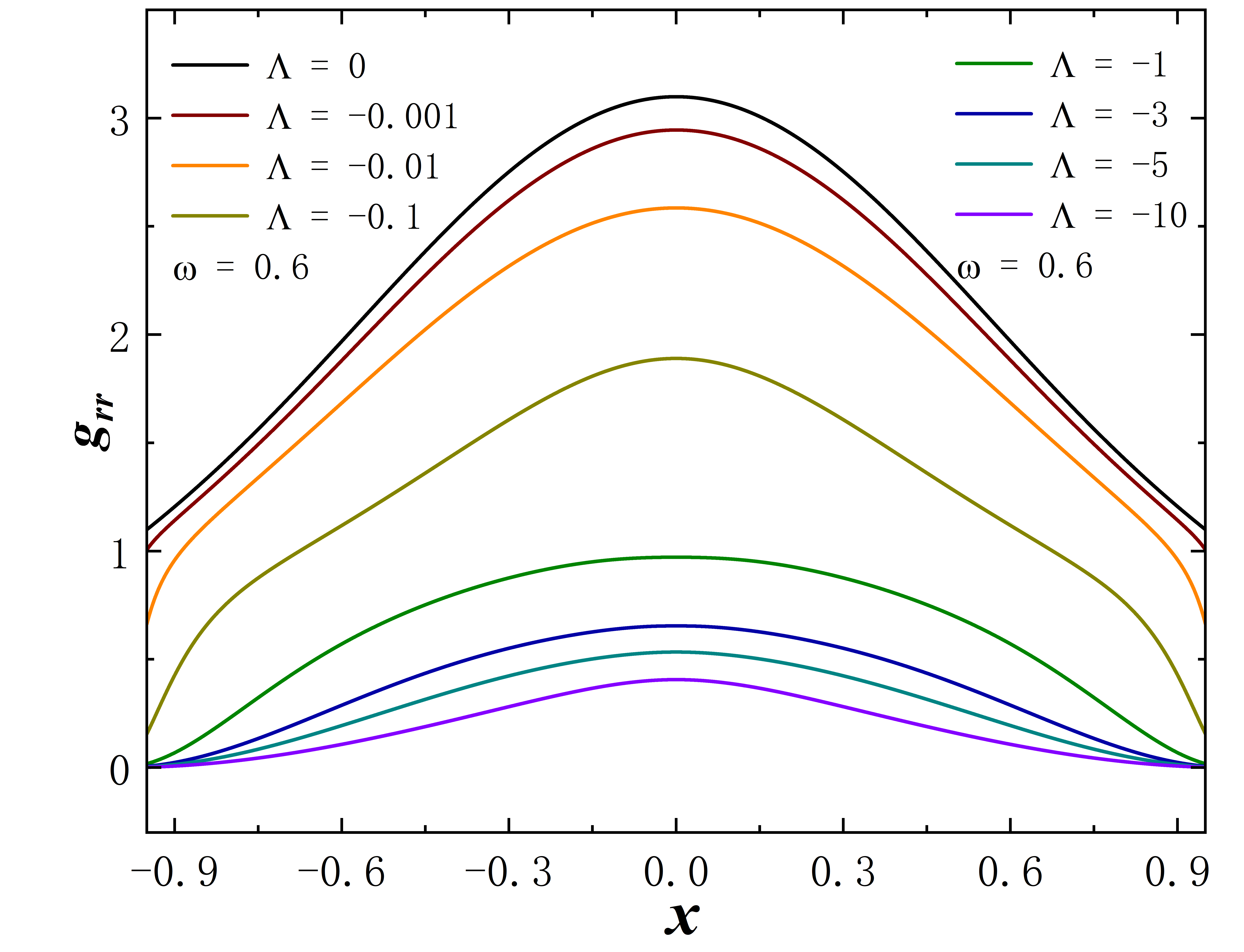}}
\subfigure{\includegraphics[width=0.45\textwidth]{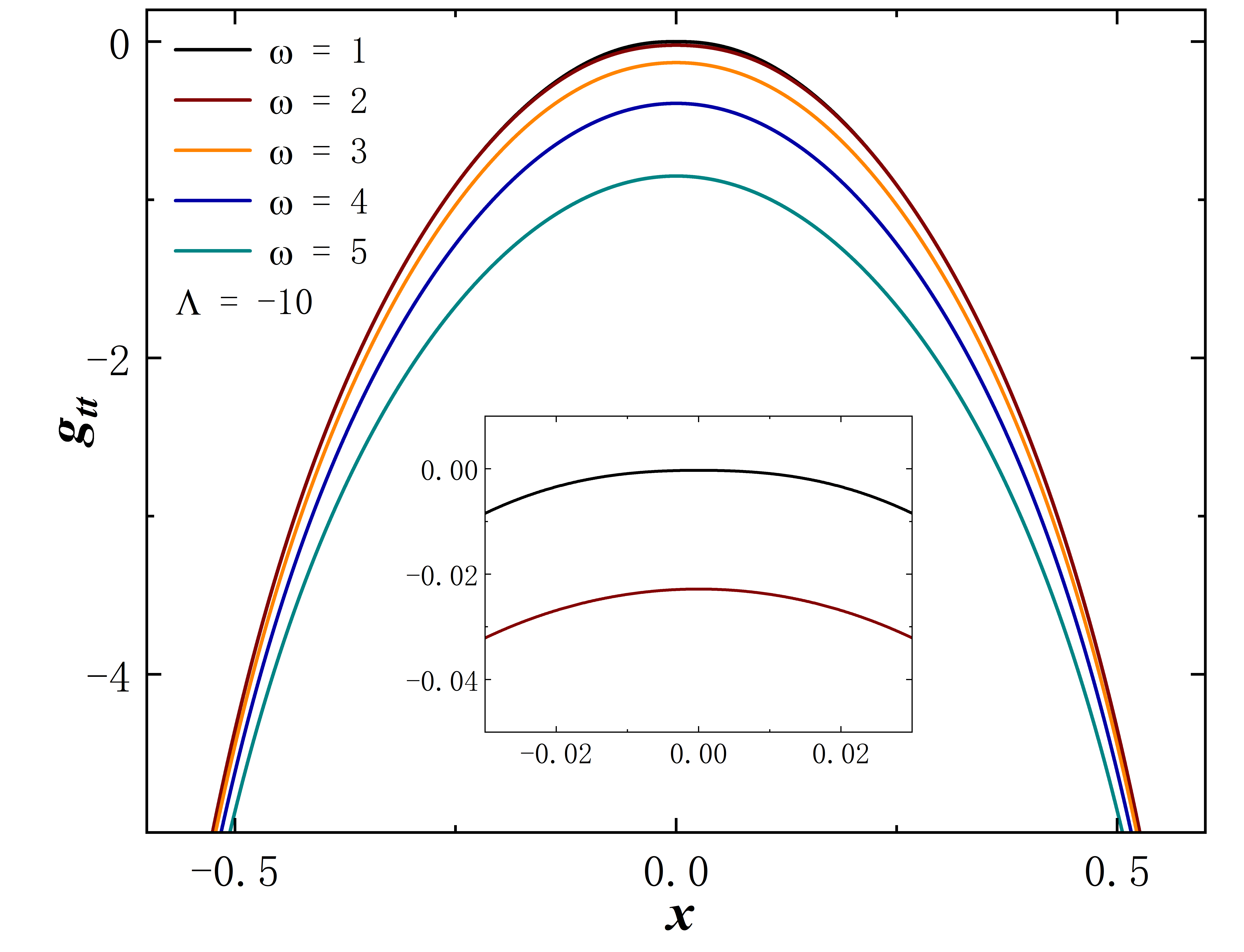}}
\subfigure{\includegraphics[width=0.45\textwidth]{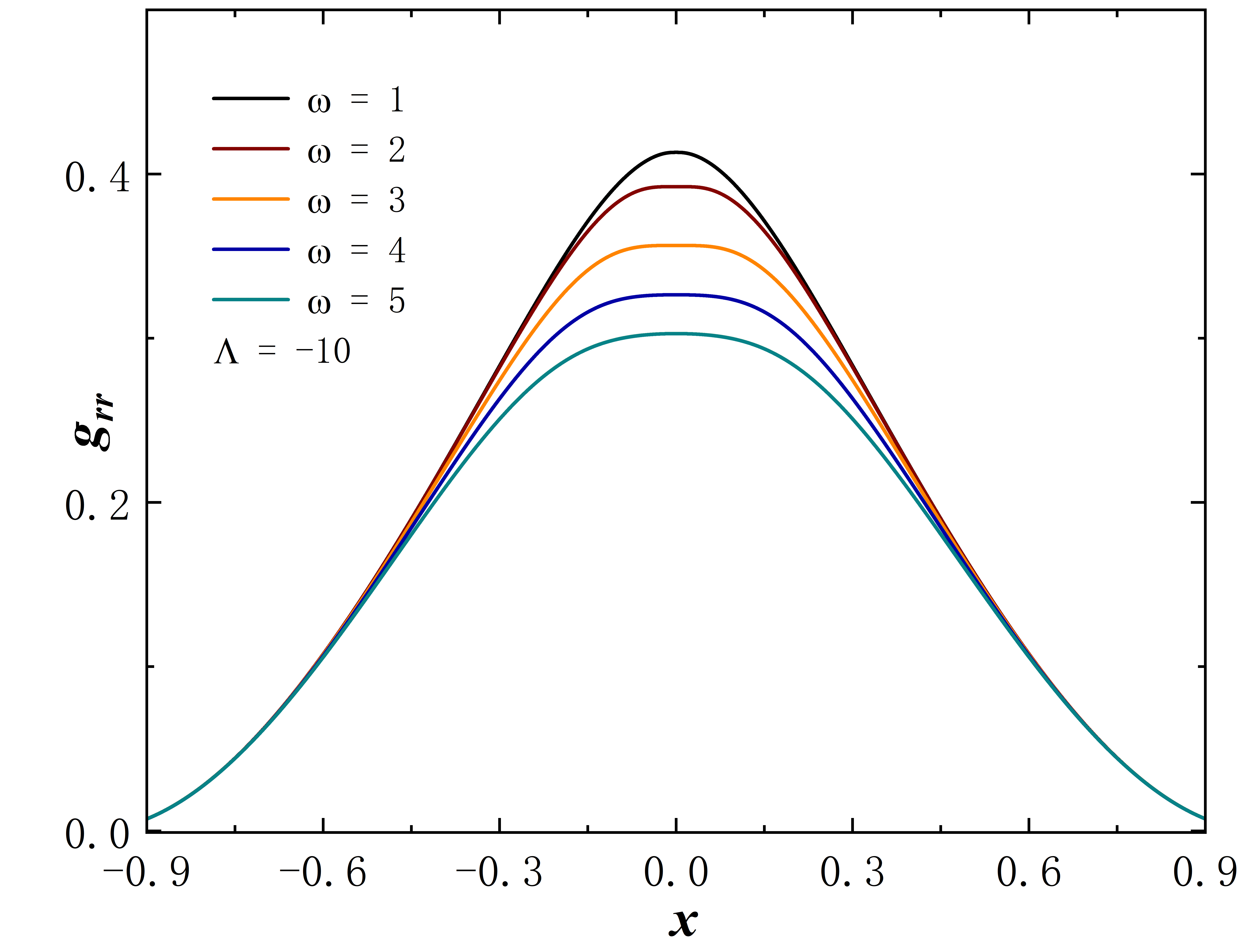}}
\end{center}
\caption{The $g_{tt}$ and $g_{rr}$ vs. the radial coordinate $x$. At the first row, the $\omega$ was fixed at 0.6 and explored different $\Lambda$ values. The fixed and varied parameters in the second row are interchanged. The throat size $r_0 = 1$.}
\label{phase3}
\end{figure}

Without loss of generality, we set $r_0 = 1$ for all subsequent numerical results. The metric functions $g_{tt}$ and $g_{rr}$ characterize the spacetime properties of the solution, as depicted in Fig. \ref{phase3}. First, we fix the frequency $\omega$ at 0.6 and explore a range of different $\Lambda$ values, and the corresponding results are depicted in the figure on the first line. When $\Lambda$ takes 0, the solution aligns with the scenario described in \cite{Ding:2023syj}. As $\Lambda$ decreases, the $g_{tt}$ value at $x = 0$ gradually approaches 0. In terms of the Eq. (8) line element, this signifies the emergence of an approximate ``event horizon''. In addition, taking the frequency to the right limit of the solution signifies the disappearance of the scalar field, causing the solution to degenerate back into the wormhole described in \cite{Blazquez-Salcedo:2020nsa}. In the second row of the figure, the cosmological constant is fixed at -10, and we investigate the variation of the metric functions with frequency $\omega$. As $\omega$ gradually decreases (approaching the left limit of the solution), the value of $g_{tt}$ at $x = 0$ once again approaches 0. The ``extreme'' behavior of the approximate ``event horizon'', resulting from parameter variations in the solution, has been investigated in \cite{Hao:2023igi, Su:2023xxk}. This phenomenon bears a resemblance to the appearance of ``one-way traversable wormholes'' due to parameter changes in the metric within the context of ``black-bounce'' scenarios \cite{Simpson:2018tsi,Lobo:2020ffi}.

To better capture the behavior of the metric function $g_{tt}$ at the throat approaching zero, we employ Tab. \ref{tab:t1} and Tab. \ref{tab:t2} to denote the value of $g_{tt}$ at $ x = 0$ under decreasing cosmological constants or frequencies. It is evident that for sufficiently small values of $\Lambda$ or $\omega$, the minimum value of $g_{tt}$ can approach $10^{-7}$ or even smaller.

    	\begin{table}[H] 
	\centering 
	\begin{tabular}{|c||c|c|c|}
\hline
		$\Lambda$ & -3 ($\omega=0.6$) & -5 ($\omega=0.6$) & -10 ($\omega=0.6$) \\
\hline
		$g_{tt}(min)$ & $-0.0012$ & $-0.00020$ & $-0.00000080$ \\
\hline
		
	\end{tabular}
 	\caption{The minimal values of $g_{tt}$ under different $\Lambda = -3, -5, -10$ with $\omega = 0.6$. The throat size $r_0 = 1$.}
	\label{tab:t1}
\end{table}

    	\begin{table}[H] 
	\centering 
	\begin{tabular}{|c||c|c|c|}
\hline
		$\omega$ & 1.5 ($\Lambda=-10$) & 1 ($\Lambda=-10$) & 0.6 ($\Lambda=-10$) \\
\hline
		$g_{tt}(min)$ & $-0.0049$ & $-0.00030$ & $-0.00000082$ \\
\hline
		
	\end{tabular}
 	\caption{The minimal values of $g_{tt}$ under different $\omega = 1.5, 1, 0.6$ with $\Lambda = -10$. The throat size $r_0 = 1$.}
	\label{tab:t2}
\end{table}

\begin{figure}
\begin{center}
\subfigure{\includegraphics[width=0.45\textwidth]{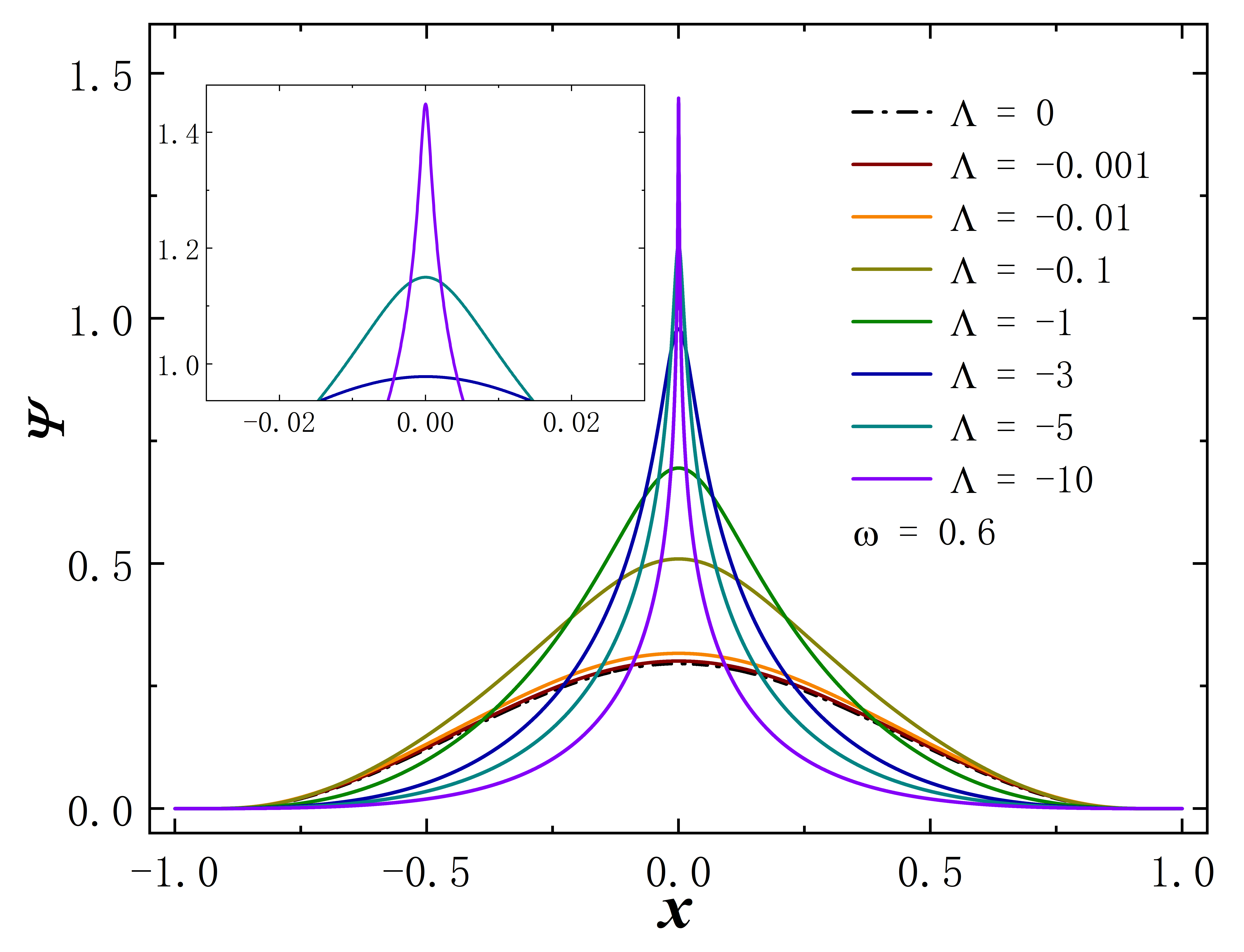}}
\subfigure{\includegraphics[width=0.45\textwidth]{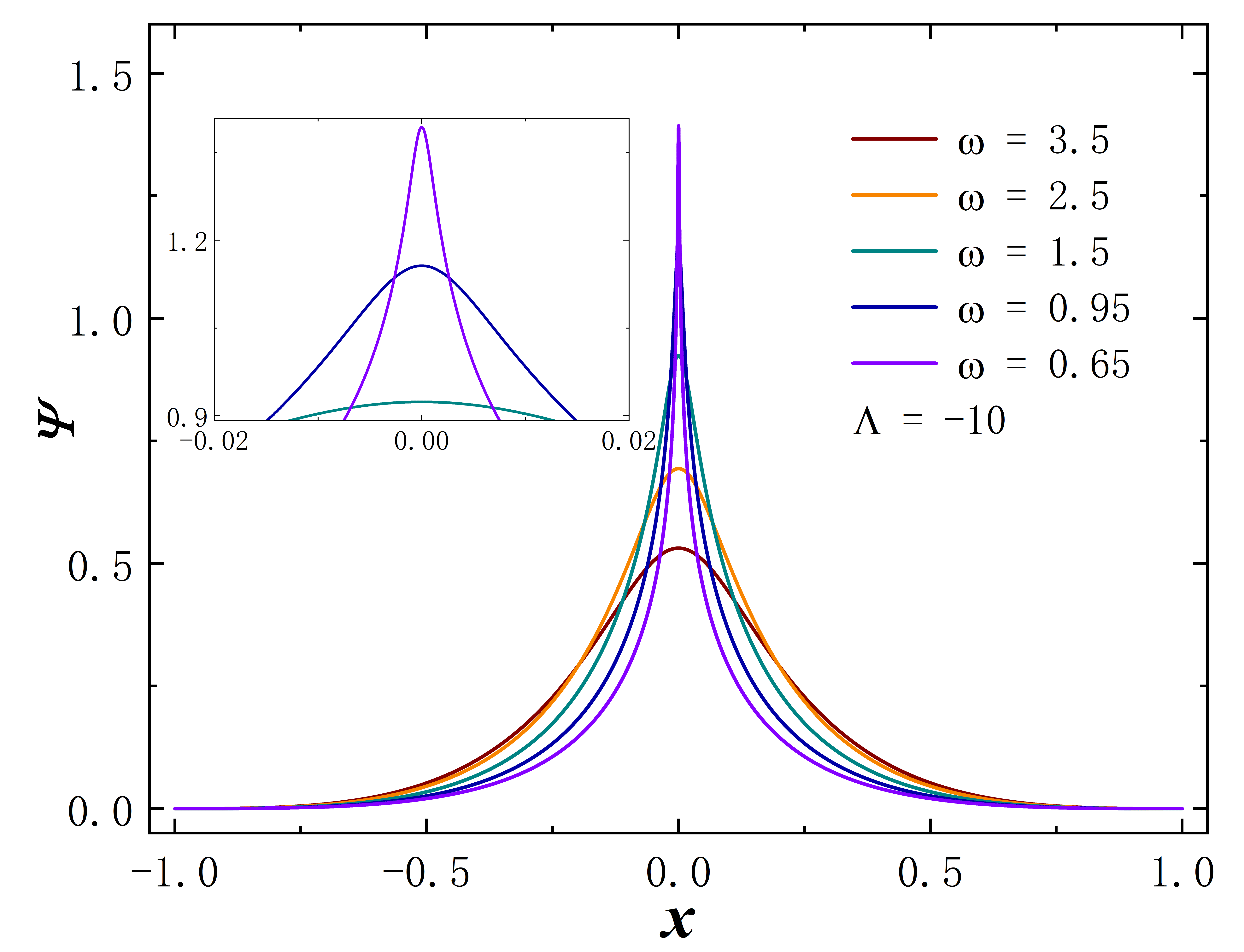}}
\end{center}
\caption{The scalar field $\psi$ vs. the radial coordinate $x$. The left and right panels depict scenarios where we fix the $\omega$ at 0.6 while varying the $\Lambda$, and vice versa, respectively. The throat size $r_0 = 1$.} 
\label{phase4}
\end{figure}

\begin{figure}
\begin{center}
\subfigure{\includegraphics[width=0.46\textwidth]{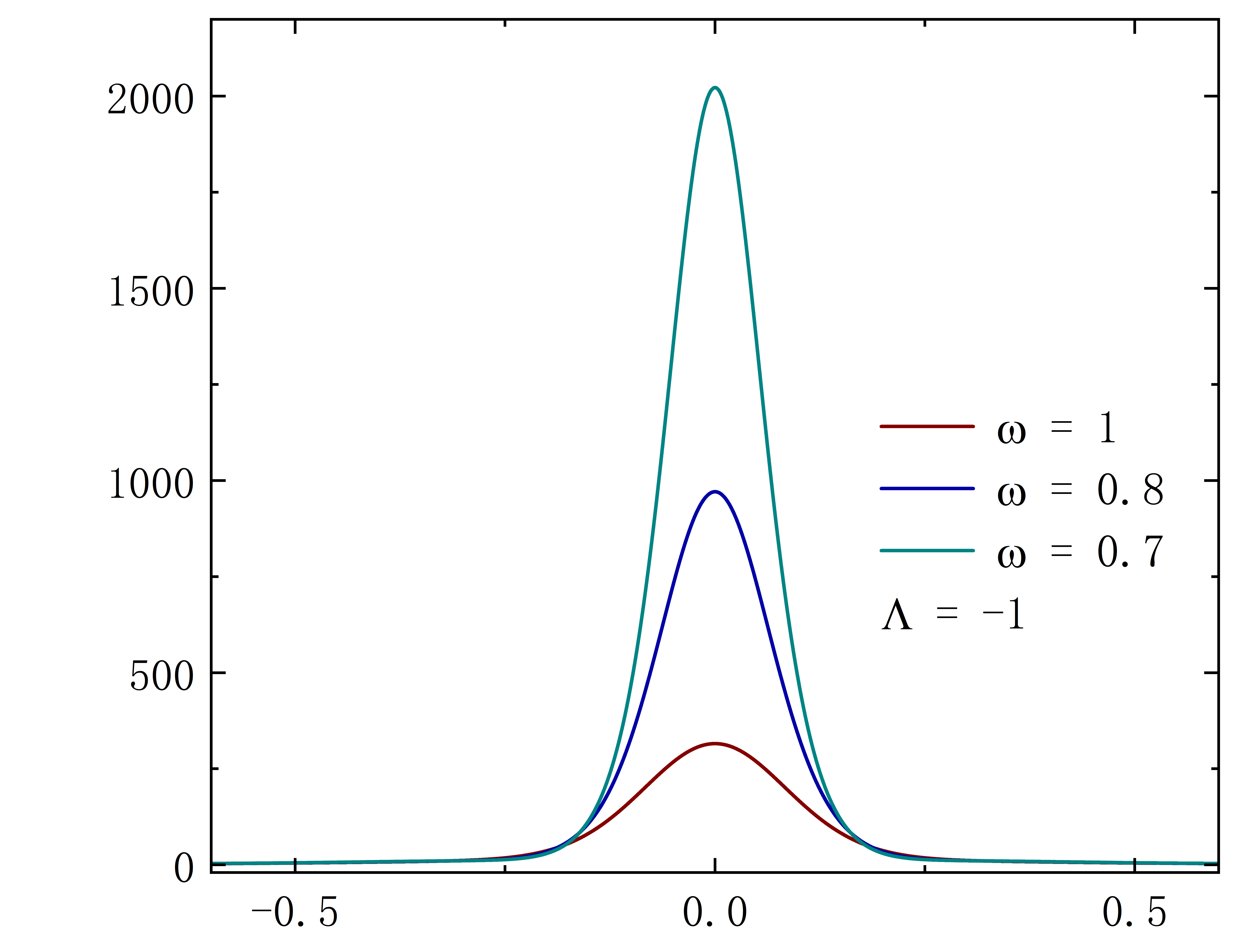}}
\subfigure{\includegraphics[width=0.46\textwidth]{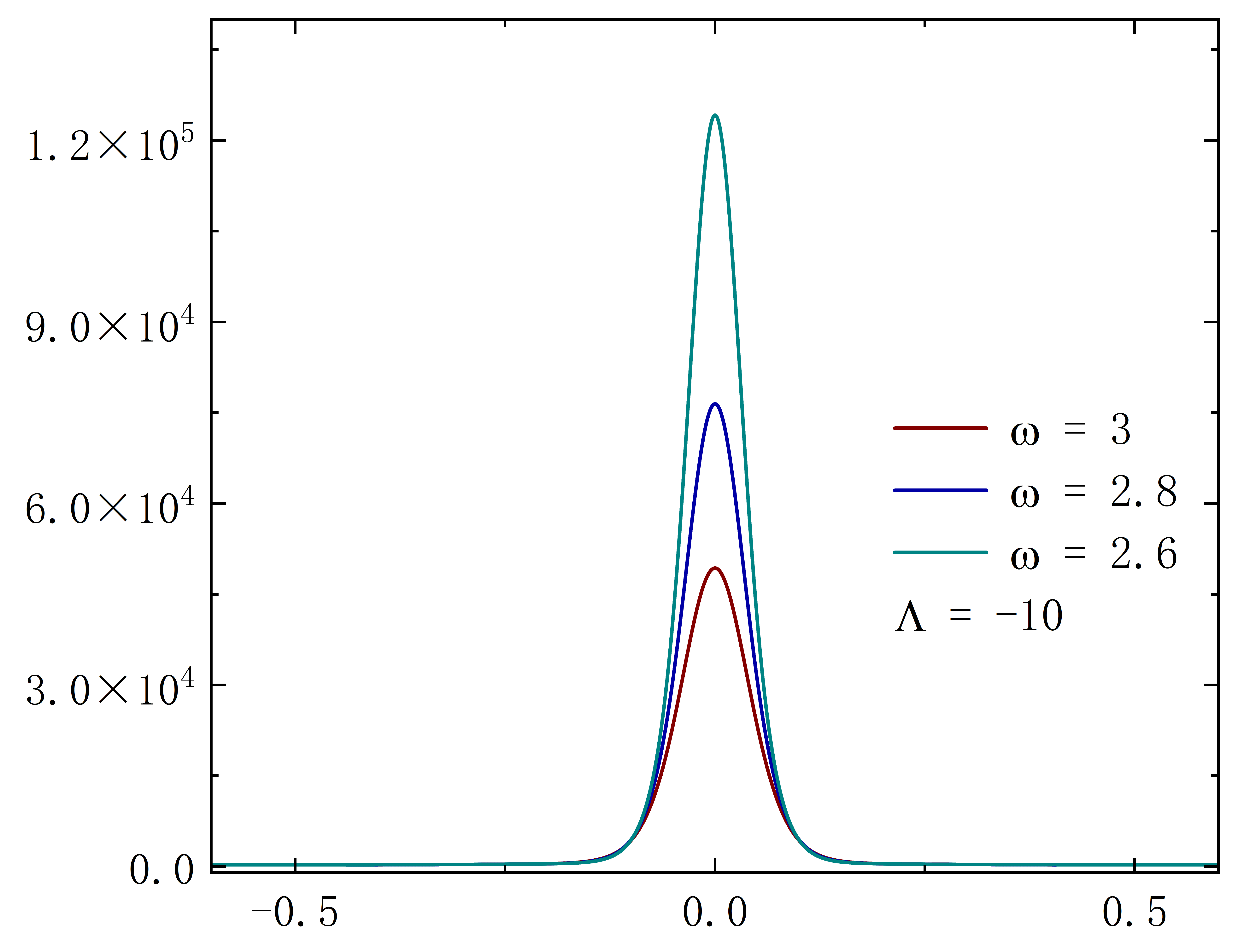}}
\end{center}
\caption{The Kretschmann scalar vs. the radial coordinate $x$. In the first row, $\Lambda$ is fixed to -1, and the $\omega$ are 1, 0.8, and 0.6 respectively. In the second row, $\Lambda$ is fixed to -10, and the $\omega$ are 3, 2.8, and 2.6 respectively. The throat size $r_0 = 1$.}
\label{phase5}
\end{figure}

Given that this traversable wormhole solution is coupled with a scalar field, investigating the properties of the scalar field becomes highly relevant. Analogous to our previous approach, we fix the frequency at 0.6 or set the cosmological constant to -10. Subsequently, we explore the effects of varying $\Lambda$ or $\omega$ on the matter field within these two scenarios Fig. \ref{phase4}. Remarkably, we observe a recurring ``extreme'' behavior. As $\Lambda$ or $\omega$ decreases, the scalar field distribution near $x = 0$ becomes highly concentrated, while the distribution in other regions of spacetime diminishes significantly. This property enables us to investigate the distribution of the Kretschmann scalar further. Fig. \ref{phase5} illustrates the Kretschmann scalar distribution when $\Lambda$ remains constant while varying $\omega$. When $\Lambda$ remains constant, decreasing $\omega$ leads to an increased value of the Kretschmann scalar, concentrating it near $x = 0$. Similarly, reducing $\Lambda$ at the same frequency $\omega$ produces a similar effect. The concentrated distribution of the matter field at the center of the wormhole, along with the nearly divergent Kretschmann scalar, suggests that within the parameter range where this ``extreme'' behavior occurs, the wormhole becomes untraversable.

\begin{figure}
\begin{center}
\subfigure{\includegraphics[width=0.45\textwidth]{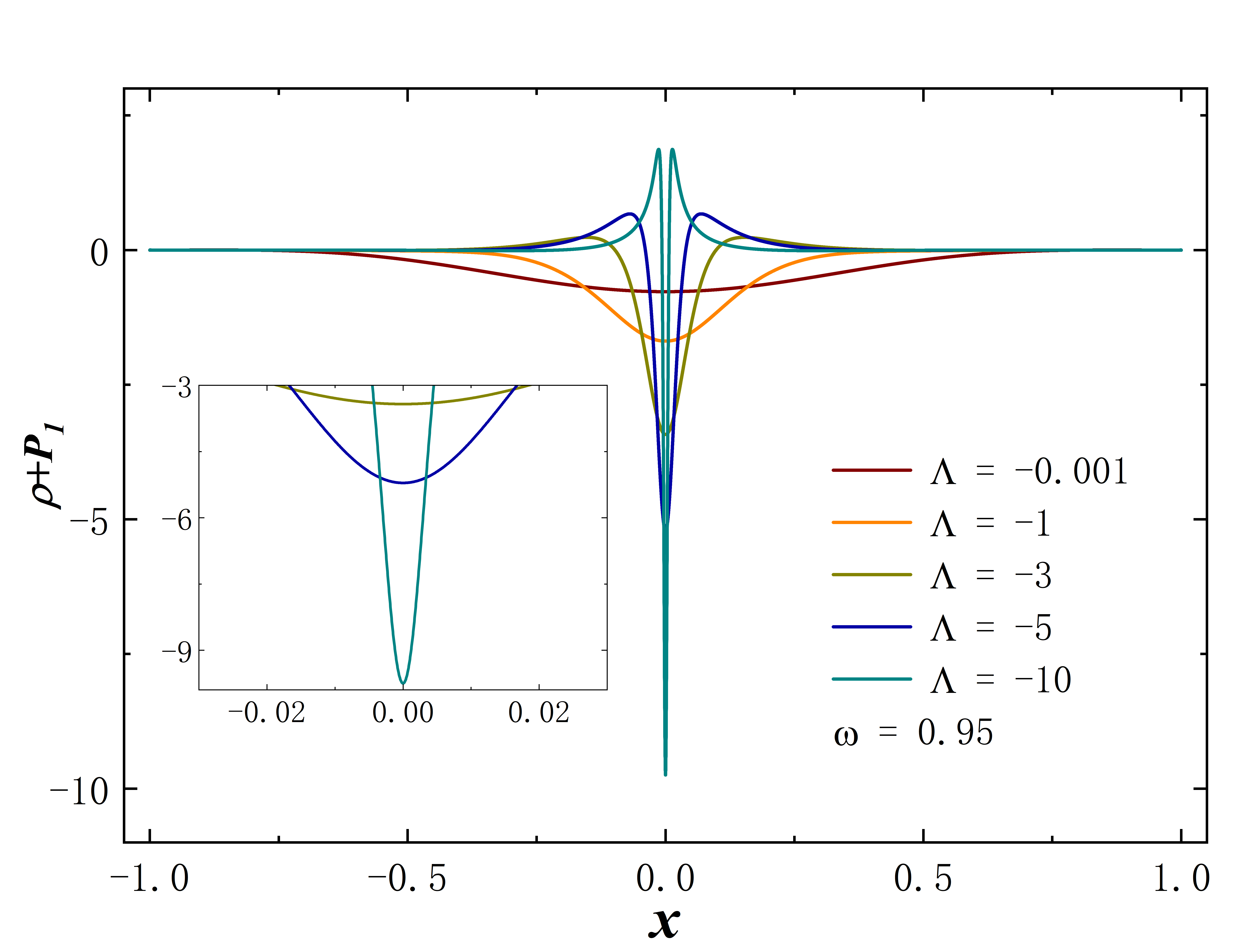}}
\subfigure{\includegraphics[width=0.45\textwidth]{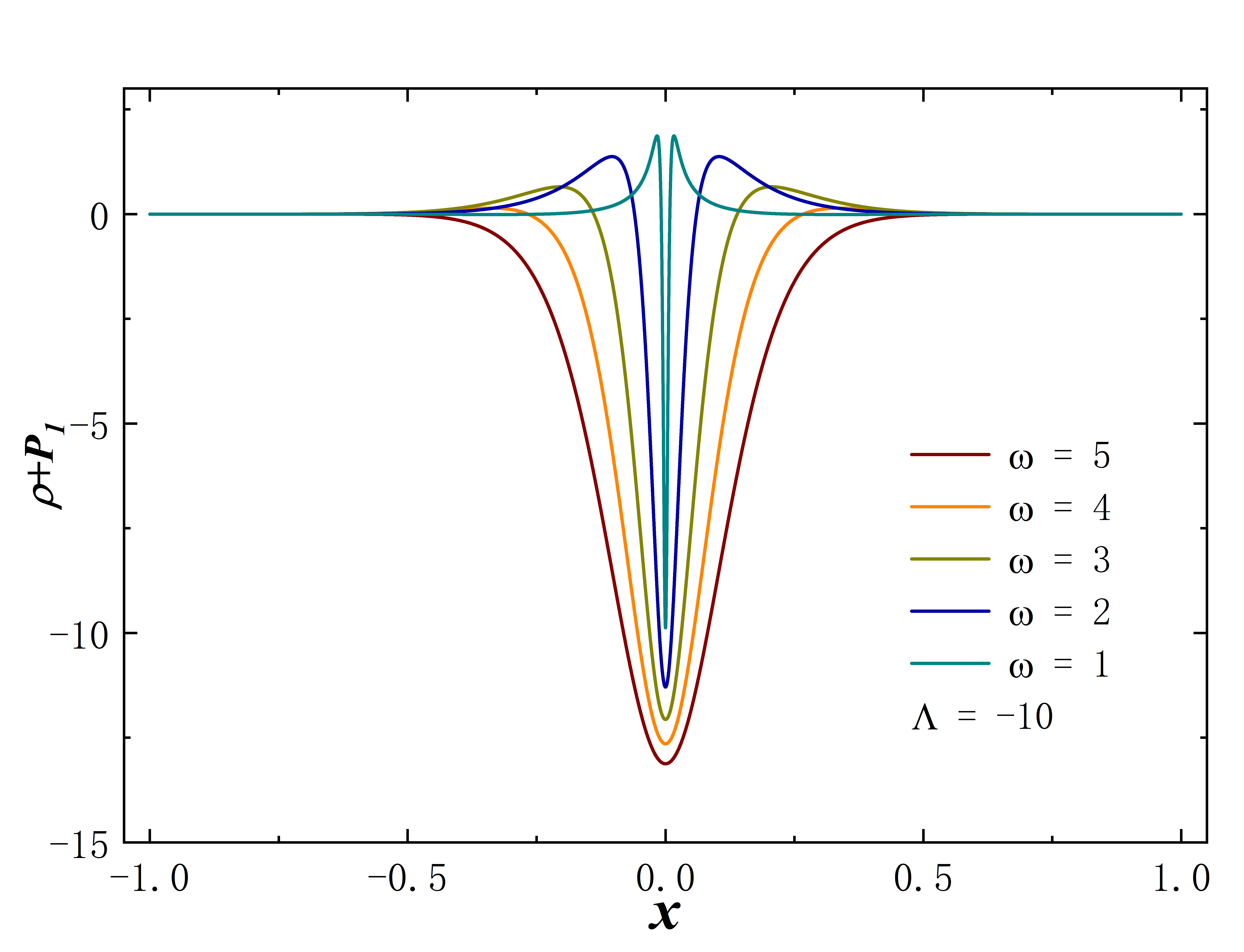}}
\end{center}
\caption{NEC violation vs. radial coordinate $x$ for different $\Lambda$ or $\omega$. The throat size $r_0 = 1$.} 
\label{phase6}
\end{figure}

Ellis wormholes devoid of matter fields violate the null energy condition (NEC) due to the presence of a phantom field. However, what happens when a scalar field is coupled? We examine the sum of energy density $\rho$ and radial pressure $p_1$ in the context of varying the cosmological constant at a fixed frequency or altering the frequency while keeping the cosmological constant Fig. \ref{phase6}. While the violation of the null energy condition (NEC) persists at the wormhole throat, the degree of violation intensifies as $\Lambda$ decreases or $\omega$ increases. However, it is noteworthy that when the cosmological constant $\Lambda$ or frequency $\omega$ becomes small (specifically when $\Lambda \le -3$ or $\omega \le 4$ in the figure), two small regions symmetrically appear on both sides of the throat. Remarkably, within these spacetime regions, the NEC remains unviolated. This effect may arise from the contribution of a scalar field with positive energy density.

\begin{figure}
\begin{center}
\subfigure{\includegraphics[width=0.45\textwidth]{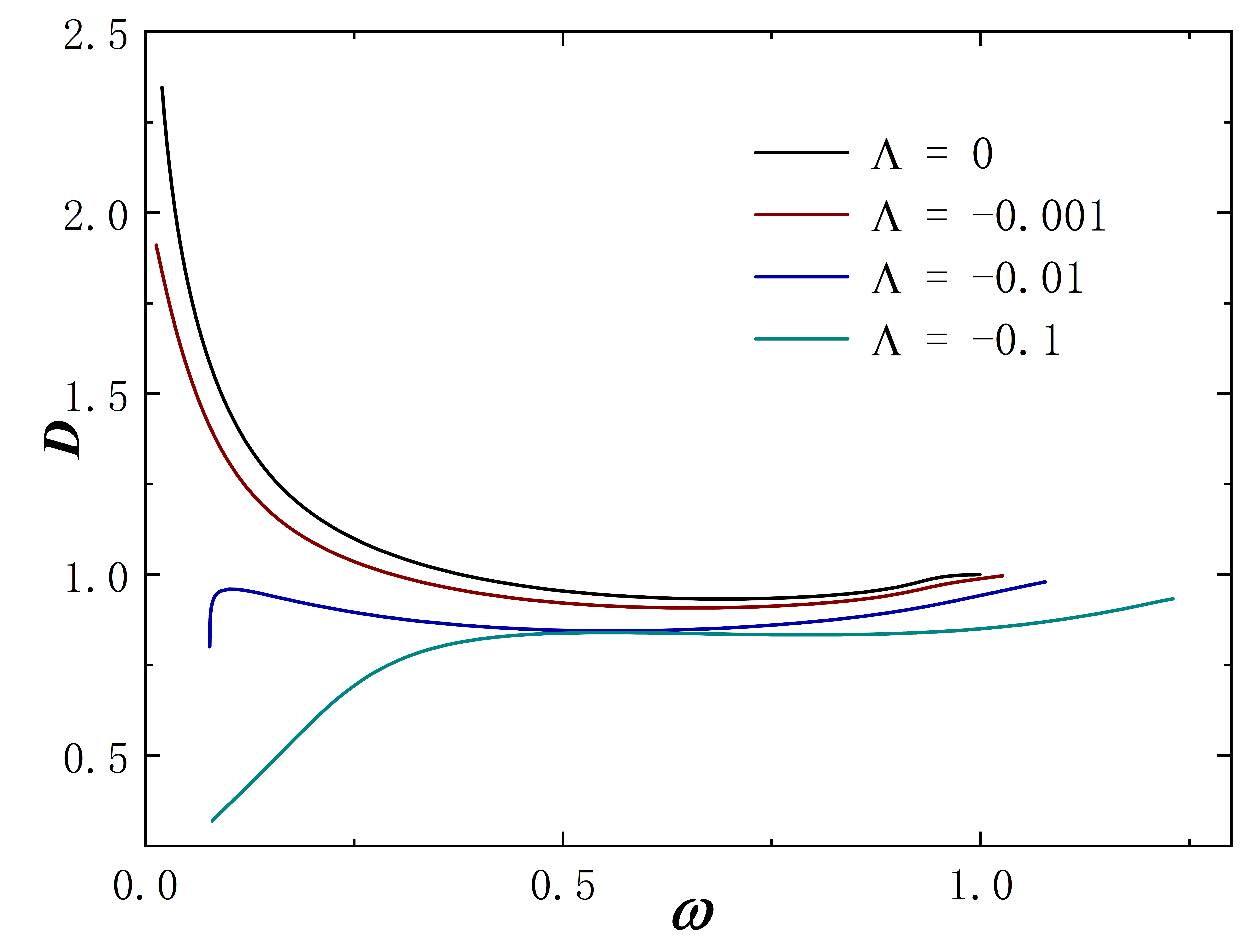}}
\subfigure{\includegraphics[width=0.45\textwidth]{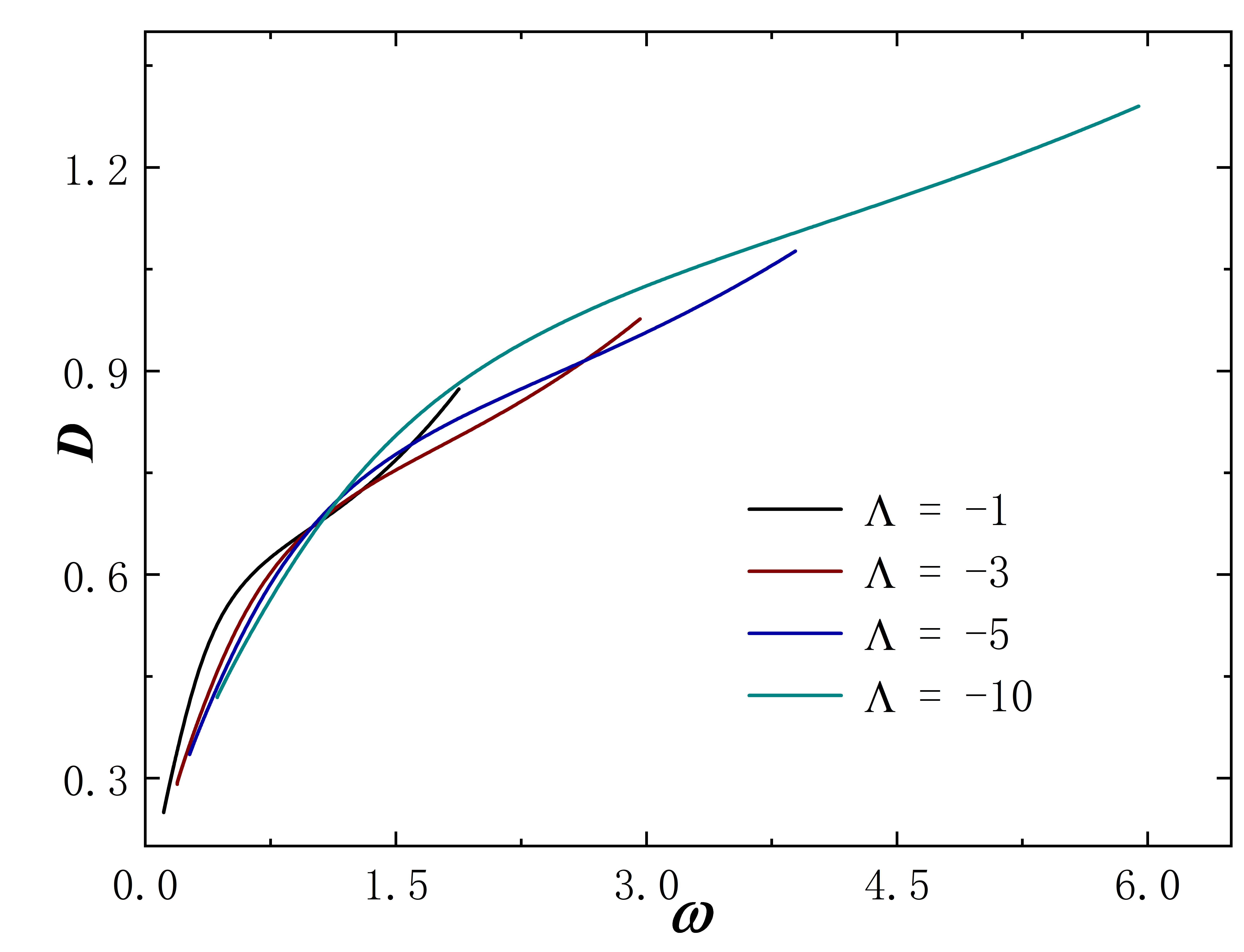}}
\end{center}
\caption{The scalar charge $\cal D$ of the phantom field as a function of the frequency $\omega$ with several values of the $\Lambda$. The throat size $r_0 = 1$.} 
\label{phase7}
\end{figure}

In the previous article, we defined the constant $\cal D$ to reflect the scalar charge of the phantom field and to test the accuracy of numerical calculations. Its value, fixed at $r_0 = 1$ as a function of frequency $\omega$, should be consistent at different positions, as shown in Fig. \ref{phase7}. Considering points at radial coordinates $x= 0.01, 0.5, 0.6, 0.9, -0.1, -0.2, -0.8$, the difference in $\cal D$ is smaller than $10^{-5}$. Analogously, if we concentrate solely on the value of $\cal D$ when the frequency is at its right limit (where the scalar field vanishes) in each scenario, the outcomes align with those observed in \cite{Blazquez-Salcedo:2020nsa}. With decreasing $\Lambda$, the constant $\cal D$ at first decreases
slightly to a minimum value and then increases as $\Lambda$ decreases further.

\subsection{The geometry properties}

Finally, we study the geometric properties of the wormhole. We can make use of a geometrical embedding diagram by fixing $t$ and $\theta$. The resulting two-dimensional spatial hypersurface of the wormhole spacetime can then be embedded in a three-dimensional Euclidean space, where the embedding diagram can be used to visualize the wormhole geometry. This technique allows us to better understand the topology and properties of the wormhole solution.

The specific method is: we begin by constructing the embeddings of planes with  $\theta = \pi/2$, and then use the cylindrical coordinates $(\rho,\varphi,z)$, the metric on this plane can be expressed by the following formula
\begin{align}
ds^2 &= \frac{p}{F N}  d r^2 + \frac{p h}{F}   d\varphi^2 \, \\
&= d \rho^2 + dz^2 + \rho^2 d \varphi^2   \,.
\end{align}
Comparing the two equations above, we then obtain the expression for $\rho$ and $z$
\begin{equation} \label{formula_embedding}
 \rho(r)= \sqrt{ \frac{p h}{F} } ,\;\;\;\;\;\;\;\;\;\;   z(r) = \pm  \int  \sqrt{ \frac{p}{F N}  -   \left( \frac{d \rho}{d r} \right)^2    }     d r \;.
\end{equation}
Here $\rho$ corresponds to the circumferential radius, which corresponds to the radius of a circle located in the equatorial plane and having a constant coordinate $r$. The function $\rho(r)$ has extreme points, where the first derivative is zero.
When the second derivative of the extreme point is greater than zero, we call this point a throat, which corresponds to a minimal surface. When the second derivative of the extreme point is less than zero, we call this point an equator, which corresponds to a maximal surface.

\begin{figure}
\begin{center}
\subfigure{\includegraphics[width=0.45\textwidth]{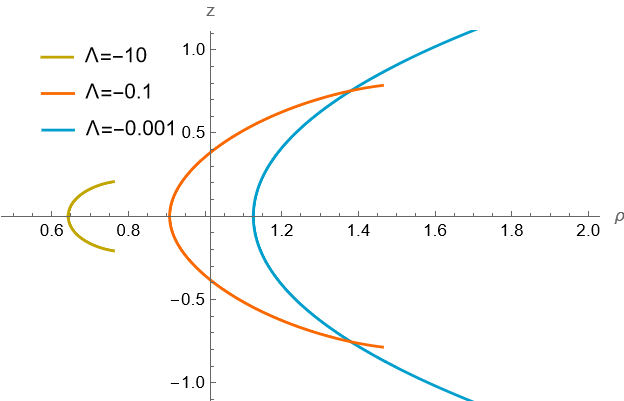}}
\subfigure{\includegraphics[width=0.45\textwidth]{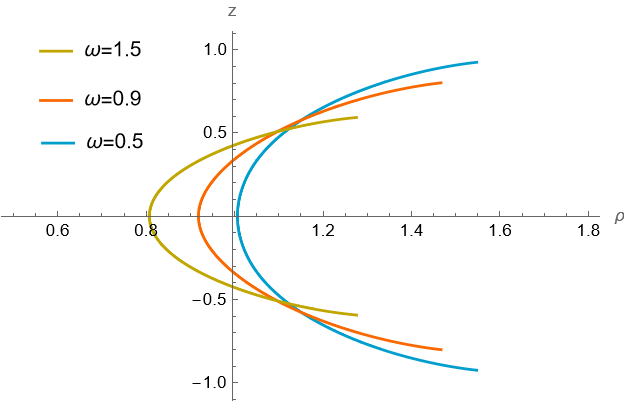}}
\subfigure{\includegraphics[width=0.9\textwidth]{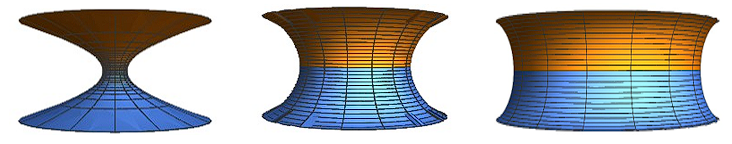}}
\subfigure{\includegraphics[width=0.88\textwidth]{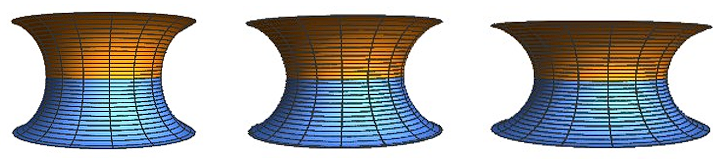}}
\end{center}
\caption{Two-dimensional view of the isometric embedding of the equatorial plane and the corresponding 3D embedding diagrams of wormhole solutions for $\Lambda = -0.001, -1, -10$ with $\omega = 0.95 $ in the first row, and $\Lambda = -1$ with $\omega = 0.5, 0.9, 1.5$ in the second row. The throat size $r_0 = 1$.} 
\label{phase8}
\end{figure}

In Fig. \ref{phase8}, we present two sets of wormhole embedding diagrams. In the first row of figures, the frequency $\omega$ is fixed at 0.95, while the cosmological constants $\Lambda$ are varied as -0.001, -1, and -10. In the second row, the $\Lambda$ remains fixed at -1, and we explore three frequencies: 0.5, 0.9, and 1.5. Wormholes exhibit inherent symmetry and consistently possess a single throat without an equatorial plane. Decreasing the cosmological constant gradually widens the wormhole's throat, whereas alterations in frequency have minimal impact on its geometric properties.

\section{CONCLUSION AND OUTLOOK}\label{sec5}

We have numerically constructed the solutions of Ellis wormholes with a scalar field in the anti-de Sitter (AdS) asymptotic spacetime. The wormhole solutions are symmetric with respect to $r \rightarrow -r$ and consequently massless. The wormholes possess a single throat which, because
of the symmetry and the radial coordinate that we have used, is located at the position $x = 0$. The solutions can exist for any value of the negative cosmological constant, in this work, we have focused on discussion on the range $-10 \le  \Lambda \le 0$. Through the computation of the Noether charge $Q$ of different cosmological constants $\Lambda$ under different $r_0$ cases, we have elucidated the properties of the matter field within this solution. Additionally, by analyzing the metric functions $g_{tt}$ and $g_{rr}$, we reveal the characteristics of spacetime. Notably, the most intriguing aspect lies in the ``extreme'' behavior observed within specific parameter ranges. At this juncture, the value of $g_{tt}$ at the throat tends toward zero, signifying the emergence of an approximate ``event horizon''. The scalar field distribution in the throat also exhibits high concentration, as reflected in the distribution of the Kretschmann scalar.

In addition, despite the persistent violation of the Null Energy Condition (NEC) at the wormhole throat, the introduction of scalar field results in NEC compliance within specific symmetry regions on both sides of the throat, subject to certain parameter ranges. Notably, wormholes exhibit geometric symmetry, featuring single throats without equatorial planes. However, variations in cosmological constants $\Lambda$ and frequencies $\omega$ lead to modifications in the wormhole embedding diagram. Finally, the substantial overlap in the values of $\cal D$ across various positions indicates that our numerical calculations exhibit minimal error.

Let us end with two related outlooks. The stability of wormholes with a phantom field has always been of concern. At present, it is known that both pure Ellis wormholes and Ellis wormholes coupled with matter fields are unstable \cite{Gonzalez:2008wd,Bronnikov:2011if,Dzhunushaliev:2014bya},  the asymptotically AdS wormholes are also unstable against radial linear perturbations \cite{Blazquez-Salcedo:2020nsa}. So what if the AdS asymptotic Ellis wormhole is coupled to the material field? We defer the investigation of stability for future work. 

While it is expected that an AdS asymptotic Ellis wormhole constructed solely from a massless phantom field remains massless, our results indicate that even when coupled with a scalar matter field, this wormhole lacks ADM mass. Notably, in \cite{Lu:2015cqa,Hao:2023kvf}, the ADM mass of black hole or wormhole spacetime transitions from positive to negative, with a solution also existing at a mass of 0. The solution obtained in this study exhibits radial symmetry, which explains the absence of ADM mass. Our future goal is to explore the possibility of finding an asymmetric AdS asymptotic wormhole solution coupled with matter.

\section*{Acknowledgements}
This work is supported by the National Key Research and Development Program of China (Grant No. 2022YFC2204101 and 2020YFC2201503) and the National Natural Science Foundation of China (Grant No. 12275110  and No. 12247101).


\begin{thebibliography}{99}

\bibitem{Einstein:1935tc}
A.~Einstein and N.~Rosen,
The Particle Problem in the General Theory of Relativity,
Phys. Rev. \textbf{48} (1935), 73-77
doi:10.1103/PhysRev.48.73

\bibitem{Kruskal:1959vx}
M.~D.~Kruskal,
Maximal extension of Schwarzschild metric,
Phys. Rev. \textbf{119} (1960), 1743-1745
doi:10.1103/PhysRev.119.1743

\bibitem{Fuller:1962zza}
R.~W.~Fuller and J.~A.~Wheeler,
Causality and Multiply Connected Space-Time,
Phys. Rev. \textbf{128} (1962), 919-929
doi:10.1103/PhysRev.128.919

\bibitem{Misner:1957mt}
C.~W.~Misner and J.~A.~Wheeler,
Classical physics as geometry: Gravitation, electromagnetism, unquantized charge, and mass as properties of curved empty space,
Annals Phys. \textbf{2} (1957), 525-603
doi:10.1016/0003-4916(57)90049-0

\bibitem{Visser:1989kh}
M.~Visser,
Traversable wormholes: Some simple examples,
Phys. Rev. D \textbf{39} (1989), 3182-3184
doi:10.1103/PhysRevD.39.3182
[arXiv:0809.0907 [gr-qc]].

\bibitem{Ellis:1973yv}
H.~G.~Ellis,
Ether flow through a drainhole - a particle model in general relativity,
J. Math. Phys. \textbf{14} (1973), 104-118
doi:10.1063/1.1666161

\bibitem{Ellis:1979bh}
H.~G.~Ellis,
THE EVOLVING, FLOWLESS DRAIN HOLE: A NONGRAVITATING PARTICLE MODEL IN GENERAL RELATIVITY THEORY,
Gen. Rel. Grav. \textbf{10} (1979), 105-123
doi:10.1007/BF00756794

\bibitem{Bronnikov:1973fh}
K.~A.~Bronnikov,
Scalar-tensor theory and scalar charge,
Acta Phys. Polon. B \textbf{4} (1973), 251-266

\bibitem{Kodama:1978dw}
T.~Kodama,
General Relativistic Nonlinear Field: A Kink Solution in a Generalized Geometry,
Phys. Rev. D \textbf{18} (1978), 3529-3534
doi:10.1103/PhysRevD.18.3529

\bibitem{Morris:1988cz}
M.~S.~Morris and K.~S.~Thorne,
Wormholes in space-time and their use for interstellar travel: A tool for teaching general relativity,
Am. J. Phys. \textbf{56} (1988), 395-412
doi:10.1119/1.15620

\bibitem{Lobo:2005us}
F.~S.~N.~Lobo,
Phys. Rev. D \textbf{71} (2005), 084011
doi:10.1103/PhysRevD.71.084011
[arXiv:gr-qc/0502099 [gr-qc]].

\bibitem{Sushkov:2005kj}
S.~V.~Sushkov,
Phys. Rev. D \textbf{71} (2005), 043520
doi:10.1103/PhysRevD.71.043520
[arXiv:gr-qc/0502084 [gr-qc]].

\bibitem{Lobo:2005yv}
F.~S.~N.~Lobo,
Phys. Rev. D \textbf{71} (2005), 124022
doi:10.1103/PhysRevD.71.124022
[arXiv:gr-qc/0506001 [gr-qc]].

\bibitem{Bronnikov:2012ch}
K.~A.~Bronnikov, R.~A.~Konoplya and A.~Zhidenko,
Phys. Rev. D \textbf{86} (2012), 024028
doi:10.1103/PhysRevD.86.024028
[arXiv:1205.2224 [gr-qc]].

\bibitem{Kleihaus:2014dla}
B.~Kleihaus and J.~Kunz,
Phys. Rev. D \textbf{90} (2014), 121503
doi:10.1103/PhysRevD.90.121503
[arXiv:1409.1503 [gr-qc]].

\bibitem{Novikov:2009vn}
D.~I.~Novikov, A.~G.~Doroshkevich, I.~D.~Novikov and A.~A.~Shatskiy,
Astron. Rep. \textbf{53} (2009), 1079-1085
doi:10.1134/S1063772909120014
[arXiv:0911.4456 [gr-qc]].

\bibitem{Bronnikov:2013coa}
K.~A.~Bronnikov, L.~N.~Lipatova, I.~D.~Novikov and A.~A.~Shatskiy,
Grav. Cosmol. \textbf{19} (2013), 269-274
doi:10.1134/S0202289313040038
[arXiv:1312.6929 [gr-qc]].

\bibitem{Cremona:2018wkj}
F.~Cremona, F.~Pirotta and L.~Pizzocchero,
Gen. Rel. Grav. \textbf{51} (2019) no.1, 19
doi:10.1007/s10714-019-2501-x
[arXiv:1805.02602 [gr-qc]].

\bibitem{Huang:2020qmn}
H.~Huang, H.~L\"u and J.~Yang,
Class. Quant. Grav. \textbf{39} (2022) no.18, 185009
doi:10.1088/1361-6382/ac8266
[arXiv:2010.00197 [gr-qc]].

\bibitem{Bronnikov:2002rn}
K.~A.~Bronnikov and S.~W.~Kim,
Possible wormholes in a brane world,
Phys. Rev. D \textbf{67} (2003), 064027
doi:10.1103/PhysRevD.67.064027
[arXiv:gr-qc/0212112 [gr-qc]].

\bibitem{Kanti:2011jz}
P.~Kanti, B.~Kleihaus and J.~Kunz,
Wormholes in Dilatonic Einstein-Gauss-Bonnet Theory
Phys. Rev. Lett. \textbf{107} (2011), 271101
doi:10.1103/PhysRevLett.107.271101
[arXiv:1108.3003 [gr-qc]].

\bibitem{Maldacena:2020sxe}
J.~Maldacena and A.~Milekhin,
Humanly traversable wormholes
Phys. Rev. D \textbf{103} (2021) no.6, 066007
doi:10.1103/PhysRevD.103.066007
[arXiv:2008.06618 [hep-th]].

\bibitem{Blazquez-Salcedo:2020czn}
J.~L.~Bl\'azquez-Salcedo, C.~Knoll and E.~Radu,
Traversable wormholes in Einstein-Dirac-Maxwell theory,
Phys. Rev. Lett. \textbf{126} (2021) no.10, 101102
doi:10.1103/PhysRevLett.126.101102
[arXiv:2010.07317 [gr-qc]].

\bibitem{Konoplya:2021hsm}
R.~A.~Konoplya and A.~Zhidenko,
Traversable Wormholes in General Relativity,
Phys. Rev. Lett. \textbf{128} (2022) no.9, 091104
doi:10.1103/PhysRevLett.128.091104
[arXiv:2106.05034 [gr-qc]].

\bibitem{Kain:2023pvp}
B.~Kain,
Einstein-Dirac-Maxwell wormholes in quantum field theory,
[arXiv:2308.00049 [gr-qc]].

\bibitem{Klinkhamer:2022rsj}
F.~R.~Klinkhamer,
Defect Wormhole: A Traversable Wormhole Without Exotic Matter,
Acta Phys. Polon. B \textbf{54} (2023) no.5, 5-A3
doi:10.5506/APhysPolB.54.5-A3
[arXiv:2301.00724 [gr-qc]].

\bibitem{Maldacena:1997re}
J.~M.~Maldacena,
Adv. Theor. Math. Phys. \textbf{2} (1998), 231-252
doi:10.4310/ATMP.1998.v2.n2.a1
[arXiv:hep-th/9711200 [hep-th]].

\bibitem{Maldacena:2013xja}
J.~Maldacena and L.~Susskind,
Fortsch. Phys. \textbf{61} (2013), 781-811
doi:10.1002/prop.201300020
[arXiv:1306.0533 [hep-th]].

\bibitem{Gao:2016bin}
P.~Gao, D.~L.~Jafferis and A.~C.~Wall,
JHEP \textbf{12} (2017), 151
doi:10.1007/JHEP12(2017)151
[arXiv:1608.05687 [hep-th]].

\bibitem{Maldacena:2017axo}
J.~Maldacena, D.~Stanford and Z.~Yang,
Fortsch. Phys. \textbf{65} (2017) no.5, 1700034
doi:10.1002/prop.201700034
[arXiv:1704.05333 [hep-th]].

\bibitem{vanBreukelen:2017dul}
R.~van Breukelen and K.~Papadodimas,
JHEP \textbf{08} (2018), 142
doi:10.1007/JHEP08(2018)142
[arXiv:1708.09370 [hep-th]].

\bibitem{Maldacena:2018lmt}
J.~Maldacena and X.~L.~Qi,
[arXiv:1804.00491 [hep-th]].

\bibitem{Dai:2020ffw}
D.~C.~Dai, D.~Minic, D.~Stojkovic and C.~Fu,
Phys. Rev. D \textbf{102} (2020) no.6, 066004
doi:10.1103/PhysRevD.102.066004
[arXiv:2002.08178 [hep-th]].

\bibitem{Bintanja:2021xfs}
S.~Bintanja, R.~Esp\'\i{}ndola, B.~Freivogel and D.~Nikolakopoulou,
JHEP \textbf{10} (2021), 173
doi:10.1007/JHEP10(2021)173
[arXiv:2102.06628 [hep-th]].

\bibitem{Kundu:2021nwp}
A.~Kundu,
Eur. Phys. J. C \textbf{82} (2022) no.5, 447
doi:10.1140/epjc/s10052-022-10376-z
[arXiv:2110.14958 [hep-th]].

\bibitem{Kain:2023ore}
B.~Kain,
Phys. Rev. Lett. \textbf{131} (2023) no.10, 101001
doi:10.1103/PhysRevLett.131.101001
[arXiv:2309.03314 [hep-th]].

\bibitem{Lemos:2003jb}
J.~P.~S.~Lemos, F.~S.~N.~Lobo and S.~Quinet de Oliveira,
Phys. Rev. D \textbf{68} (2003), 064004
doi:10.1103/PhysRevD.68.064004
[arXiv:gr-qc/0302049 [gr-qc]].

\bibitem{Lemos:2004vs}
J.~P.~S.~Lemos and F.~S.~N.~Lobo,
Phys. Rev. D \textbf{69} (2004), 104007
doi:10.1103/PhysRevD.69.104007
[arXiv:gr-qc/0402099 [gr-qc]].

\bibitem{Maeda:2008nz}
H.~Maeda and M.~Nozawa,
Phys. Rev. D \textbf{78} (2008), 024005
doi:10.1103/PhysRevD.78.024005
[arXiv:0803.1704 [gr-qc]].

\bibitem{Maeda:2012fr}
H.~Maeda,
Phys. Rev. D \textbf{86} (2012), 044016
doi:10.1103/PhysRevD.86.044016
[arXiv:1204.4472 [gr-qc]].

\bibitem{Wu:2022gpm}
T.~Wu,
Phys. Rev. D \textbf{108} (2023) no.4, 044001
doi:10.1103/PhysRevD.108.044001
[arXiv:2209.02278 [gr-qc]].

\bibitem{Nozawa:2020gzz}
M.~Nozawa,
Phys. Rev. D \textbf{103} (2021) no.2, 024005
doi:10.1103/PhysRevD.103.024005
[arXiv:2010.07561 [gr-qc]].

\bibitem{Anabalon:2018rzq}
A.~Anabal\'on and J.~Oliva,
JHEP \textbf{04} (2019), 106
doi:10.1007/JHEP04(2019)106
[arXiv:1811.03497 [hep-th]].

\bibitem{Korolev:2014hwa}
R.~V.~Korolev and S.~V.~Sushkov,
Phys. Rev. D \textbf{90} (2014), 124025
doi:10.1103/PhysRevD.90.124025
[arXiv:1408.1235 [gr-qc]].

\bibitem{Franciolini:2018aad}
G.~Franciolini, L.~Hui, R.~Penco, L.~Santoni and E.~Trincherini,
JHEP \textbf{01} (2019), 221
doi:10.1007/JHEP01(2019)221
[arXiv:1811.05481 [hep-th]].

\bibitem{Chatzifotis:2020oqr}
N.~Chatzifotis, G.~Koutsoumbas and E.~Papantonopoulos,
Phys. Rev. D \textbf{104} (2021) no.2, 024039
doi:10.1103/PhysRevD.104.024039
[arXiv:2011.08770 [gr-qc]].

\bibitem{Blazquez-Salcedo:2020nsa}
J.~L.~Bl\'azquez-Salcedo, X.~Y.~Chew, J.~Kunz and D.~H.~Yeom,
Eur. Phys. J. C \textbf{81} (2021) no.9, 858
doi:10.1140/epjc/s10052-021-09645-0
[arXiv:2012.06213 [gr-qc]].

\bibitem{Wheeler:1955zz}
J.~A.~Wheeler,
Phys. Rev. \textbf{97} (1955), 511-536
doi:10.1103/PhysRev.97.511

\bibitem{Power:1957zz}
E.~A.~Power and J.~A.~Wheeler,
Rev. Mod. Phys. \textbf{29} (1957), 480-495
doi:10.1103/RevModPhys.29.480

\bibitem{Kaup:1968zz}
D.~J.~Kaup,
Phys. Rev. \textbf{172} (1968), 1331-1342
doi:10.1103/PhysRev.172.1331

\bibitem{Ruffini:1969qy}
R.~Ruffini and S.~Bonazzola,
Phys. Rev. \textbf{187} (1969), 1767-1783
doi:10.1103/PhysRev.187.1767

\bibitem{Schunck:1996he}
F.~E.~Schunck and E.~W.~Mielke,
Phys. Lett. A \textbf{249} (1998), 389-394
doi:10.1016/S0375-9601(98)00778-6

\bibitem{Yoshida:1997qf}
S.~Yoshida and Y.~Eriguchi,
Phys. Rev. D \textbf{56} (1997), 762-771
doi:10.1103/PhysRevD.56.762

\bibitem{Bernal:2009zy}
A.~Bernal, J.~Barranco, D.~Alic and C.~Palenzuela,
Phys. Rev. D \textbf{81} (2010), 044031
doi:10.1103/PhysRevD.81.044031
[arXiv:0908.2435 [gr-qc]].

\bibitem{Collodel:2017biu}
L.~G.~Collodel, B.~Kleihaus and J.~Kunz,
Phys. Rev. D \textbf{96} (2017) no.8, 084066
doi:10.1103/PhysRevD.96.084066
[arXiv:1708.02057 [gr-qc]].

\bibitem{Wang:2018xhw}
Y.~Q.~Wang, Y.~X.~Liu and S.~W.~Wei,
Phys. Rev. D \textbf{99} (2019) no.6, 064036
doi:10.1103/PhysRevD.99.064036
[arXiv:1811.08795 [gr-qc]].

\bibitem{Herdeiro:2021mol}
C.~A.~R.~Herdeiro, J.~Kunz, I.~Perapechka, E.~Radu and Y.~Shnir,
Phys. Rev. D \textbf{103} (2021) no.6, 065009
doi:10.1103/PhysRevD.103.065009
[arXiv:2101.06442 [gr-qc]].

\bibitem{Astefanesei:2003qy}
D.~Astefanesei and E.~Radu,
Nucl. Phys. B \textbf{665} (2003), 594-622
doi:10.1016/S0550-3213(03)00482-6
[arXiv:gr-qc/0309131 [gr-qc]].

\bibitem{Buchel:2013uba}
A.~Buchel, S.~L.~Liebling and L.~Lehner,
Phys. Rev. D \textbf{87} (2013) no.12, 123006
doi:10.1103/PhysRevD.87.123006
[arXiv:1304.4166 [gr-qc]].

\bibitem{Maliborski:2013ula}
M.~Maliborski and A.~Rostworowski,
[arXiv:1307.2875 [gr-qc]].

\bibitem{Fodor:2015eia}
G.~Fodor, P.~Forg\'acs and P.~Grandcl\'ement,
Phys. Rev. D \textbf{92} (2015) no.2, 025036
doi:10.1103/PhysRevD.92.025036
[arXiv:1503.07746 [gr-qc]].

\bibitem{Brihaye:2013hx}
Y.~Brihaye, B.~Hartmann and S.~Tojiev,
Class. Quant. Grav. \textbf{30} (2013), 115009
doi:10.1088/0264-9381/30/11/115009
[arXiv:1301.2452 [hep-th]].

\bibitem{Liu:2020uaz}
H.~S.~Liu, H.~Lu and Y.~Pang,
Phys. Rev. D \textbf{102} (2020) no.12, 126008
doi:10.1103/PhysRevD.102.126008
[arXiv:2007.15017 [hep-th]].

\bibitem{Dzhunushaliev:2014bya}
V.~Dzhunushaliev, V.~Folomeev, C.~Hoffmann, B.~Kleihaus and J.~Kunz,
Phys. Rev. D \textbf{90} (2014) no.12, 124038
doi:10.1103/PhysRevD.90.124038
[arXiv:1409.6978 [gr-qc]].

\bibitem{Hoffmann:2017jfs}
C.~Hoffmann, T.~Ioannidou, S.~Kahlen, B.~Kleihaus and J.~Kunz,
Phys. Rev. D \textbf{95} (2017) no.8, 084010
doi:10.1103/PhysRevD.95.084010
[arXiv:1703.03344 [gr-qc]].

\bibitem{Yue:2023ela}
Y.~Yue, P.~B.~Ding and Y.~Q.~Wang,
Eur. Phys. J. C \textbf{83} (2023) no.8, 732
doi:10.1140/epjc/s10052-023-11914-z
[arXiv:2305.04496 [gr-qc]].

\bibitem{Ding:2023syj}
P.~B.~Ding, T.~X.~Ma and Y.~Q.~Wang,
[arXiv:2305.19819 [gr-qc]].

\bibitem{Hao:2023igi}
C.~H.~Hao, S.~X.~Sun, L.~X.~Huang, R.~Zhang, X.~Su and Y.~Q.~Wang,
[arXiv:2309.16379 [gr-qc]].

\bibitem{Su:2023xxk}
X.~Su, C.~H.~Hao, J.~R.~Ren and Y.~Q.~Wang,
[arXiv:2311.17557 [gr-qc]].

\bibitem{Simpson:2018tsi}
A.~Simpson and M.~Visser,
Black-bounce to traversable wormhole,
JCAP \textbf{02} (2019), 042
doi:10.1088/1475-7516/2019/02/042
[arXiv:1812.07114 [gr-qc]].

\bibitem{Lobo:2020ffi}
F.~S.~N.~Lobo, M.~E.~Rodrigues, M.~V.~de Sousa Silva, A.~Simpson and M.~Visser,
Novel black-bounce spacetimes: wormholes, regularity, energy conditions, and causal structure,
Phys. Rev. D \textbf{103} (2021) no.8, 084052
doi:10.1103/PhysRevD.103.084052
[arXiv:2009.12057 [gr-qc]].

\bibitem{Gonzalez:2008wd}
J.~A.~Gonzalez, F.~S.~Guzman and O.~Sarbach,
Class. Quant. Grav. \textbf{26} (2009), 015010
doi:10.1088/0264-9381/26/1/015010
[arXiv:0806.0608 [gr-qc]].

\bibitem{Bronnikov:2011if}
K.~A.~Bronnikov, J.~C.~Fabris and A.~Zhidenko,
Eur. Phys. J. C \textbf{71} (2011), 1791
doi:10.1140/epjc/s10052-011-1791-2
[arXiv:1109.6576 [gr-qc]].

\bibitem{Lu:2015cqa}
H.~Lu, A.~Perkins, C.~N.~Pope and K.~S.~Stelle,
Phys. Rev. Lett. \textbf{114} (2015) no.17, 171601
doi:10.1103/PhysRevLett.114.171601
[arXiv:1502.01028 [hep-th]].

\bibitem{Hao:2023kvf}
C.~H.~Hao, L.~X.~Huang, X.~Su and Y.~Q.~Wang,
[arXiv:2312.03800 [gr-qc]].







\end{thebibliography}
\end{document}